\documentclass{sig-alternate-10pt} 

\usepackage[T1]{fontenc}
\usepackage[latin1]{inputenc}

\usepackage{german}
\usepackage[tight]{subfigure}
\graphicspath{{figs/}}

\usepackage{url}

\begin{document}
\clubpenalty=10000
\widowpenalty = 10000

\title{Backscatter from the Data Plane -- Threats to Stability and Security in Information-Centric Networking}
\subtitle{Initial Version 1 -- May 22, 2012 \\ Version 2 -- Updated on \today}

\numberofauthors{3}
\author{
\alignauthor Matthias W{\"a}hlisch\\
       \affaddr{Freie Universit"at Berlin}\\
       \affaddr{Inst. f"ur Informatik, Takustr. 9}\\
       \affaddr{D--14195 Berlin, Germany}\\
       \email{waehlisch@ieee.org}
\alignauthor Thomas C. Schmidt\\
       \affaddr{HAW Hamburg, Informatik}\\
       \affaddr{Berliner Tor 7}\\
       \affaddr{D--20099 Hamburg, Germany}\\
       \email{t.schmidt@ieee.org}
\alignauthor Markus Vahlenkamp\\
       \affaddr{HAW Hamburg, Informatik}\\
       \affaddr{Berliner Tor 7}\\
       \affaddr{D--20099 Hamburg, Germany}\\
       \email{markus@vahlenkamp.net}
}
\maketitle

\begin{abstract}
Information-centric networking proposals attract much attention in the ongoing search for a future communication paradigm of the Internet. Replacing the host-to-host connectivity by a data-oriented  publish/subscribe service eases content distribution and authentication by concept, while eliminating threats from unwanted traffic at an end host as are common in today's Internet. However, current approaches to content routing heavily rely on data-driven protocol events and thereby introduce a strong coupling of the control to the data plane in the underlying routing infrastructure. In this paper, threats to the stability and security of the content distribution system are analyzed in theory and practical experiments. We derive relations between state resources and the performance of routers and demonstrate how this coupling can be misused in practice. We discuss new attack vectors present in its current state of development, as well as possibilities and limitations to mitigate them.  
\end{abstract}

\protect
\keywords{Performance, Forwarding Plane, Security, Distributed Denial of Service (DDoS), Vulnerability, Content-centric Routing, Resource Exhaustion, Jamming, Named-data Networking}
\vspace{-0.75\baselineskip}
\section{Introduction}

One major dedication of today's Internet is the global distribution of content in huge amounts. Content distribution networks (CDNs) facilitate an efficient, wide-area replication of static data for selected content pro\-vi\-ders, whereas the end-to-end design of TCP/IP does not foresee implicit replication and in-network storage. There is no openly available network standard for the asynchronous, global replication of popular content in the current Internet.

Inspired by the use case of widely deployed CDNs, current research has created the concepts of {\em Content} or {\em Information-Centric Networks}. We will use the acronym ICN to subsume the different approaches throughout this article.  ICN abandons the current Internet model of connecting end nodes. Instead, it follows a data-oriented  publish/subscribe para\-digm that facilitates content distribution and authentication by concept. In ICN, consumers shall retrieve content by name directly from a network that provides storage, caching, content-based rendezvous, and searching at times.

Several proposals have been presented in recent years \cite{adiko-sind-12}, among them TRIAD \cite{gc-acrsi-01}, DONA \cite{kccek-dona-07}, NDN \cite{zebjt-ndnp-10,jstp-nnc-09}, PSIRP \cite{jzran-llsps-09}, and NetInf \cite{a-snad-10}, which differ in several design choices. As we are interested in the stability and security of ICN infrastructures, we will concentrate on the aspects of routing and forwarding.

Essentially two approaches to routing exist in current ICN proposals, an evolutionary path that resolves names to locators and routes on IP (or a related location scheme), and  `clean slate' concepts that route directly on content identities. NetInf extends the current Internet by a resolution service that maps content names to topological IDs like IP addresses, but alternatively supports name-based routing. TRIAD, DONA and NDN perform content retrieval by routing on names. Route responses and the data itself are then forwarded along reverse paths (RPF), either by using IP as a lower layer, or without IP but by dedicated RPF states. PSIRP publishes content objects to a resolution system that incloses full knowledge of the network topology. Requesters trigger the mapping system to generate forwarding identifiers, i.e., Bloom filters of aggregated link IDs, that are used to forward of data.

All solutions operate on the content itself, and force the network infrastructure into a content awareness. A mapping service is not only required to resolve {\em file} names to source locations, but must answer a request by advising a nearby replica, the existence of which it learned from the data distribution system. Content routers need to rely on (often aggregated) names in its interface tables  and -- for RPF-based forwarding schemes -- a reverse state for every data unit. This control information is highly dynamic and requires regular updates from the data plane. The ICN paradigm thereby opens up the control plane to continuous modifications from the data plane. This is in contrast to the current Internet, where DNS and routing states remain unaltered by data-driven events such as file names or cache locations.

In this paper, we study the effects of control plane behaviour under various data conditions. We are in particular interested in threats to the stability and security of the ICN infrastructure, whose impacts we evaluate in a theoretical analysis and experimental trials. Experiments are performed in test networks running PARC's CCNx software. We want to stress, though, that our tests only attribute for the core concepts  of content routing and do not evaluate implementation properties of the CCNx prototype.  Following the basic insights gained from theoretical and practical analysis, we contribute a collection of new attacks that ground on this transfer from data to control states.

The remainder of this paper is organized as follows: The specific problems in protecting  
the ICN infrastructure  are stated in \S~\ref{sec:ps} along with related work on
ICN security. We theoretically analyze basic threats to stability in  \S~\ref{sec:basics} and discuss related implications. Based on practical experiments, threatening scenarios and their effects on the routing system are demonstrated in \S~\ref{sec:evaluation}. These general insights lead to concrete attack
scenarios in \S~\ref{sec:attacks}. The paper concludes with a discussion
in \S~\ref{sec:discussion}.

\section{Why ICN is Challenged by Design}\label{sec:ps}

\subsection{Problem Statement}

Information-centric networking introduces two kinds of states into the network infrastructure, (1) content publications or announcements, and (2) subscriptions or requests for asynchronous content access. Publication states are announced within the request routing system or stored in some name resolution service. Subscriptions are memorized hop by hop in reverse path forwarding states, in the name system, or at communicating end nodes in IP-based systems.\footnote{LIPSIN, as an exception, keeps record of subscriptions in the mapping/rendezvous system to construct source routing filters that generate distribution trees.} In addition, in-network caching requires management of replicas at content routers. These statefull operations raise the following issues:

\begin{description}
\item[Resource Exhaustion:] Infrastructural entities need to offer accumulating resources like memory and processing power for provisioning, maintaining and exchanging content states. They are therefore threatened by resource exhaustion due to misuse or uncontrolled load.

\item[State Decorrelation:] The asynchronous nature of publish/subscribe content delivery places the enhanced burden of assuring consistency among distributed data states. Data states that require correlation are situated in distributed mapping systems, which also need to consistently reflect actual content placements, and in forwarding states at routers that  define the paths hop-by-hop from a supplier to the requester. Failures in state coherence may lead to service disruptions or unwanted traffic flows.

\item[Path \& Name Infiltration:] The infrastructure relies on the integrity and correctness of content routing and is therefore threatened by poisonous injections of paths and names, in particular. The replicative ICN environment distributes content copies to many, commonly untrusted locations and thereby makes it particularly hard to authenticate valid origins of state insertion requests.

\item[Cache Pollution:] The conceptual value and the performance of the ICN infrastructure is built upon receiver-driven caching. It is therefore vulnerable to all operations that spoil the cache relevance.

\item[Cryptographic Breaches:] Publishers, who massively sign long-lived content, offer time and data to an attacker for compromising cryptographic credentials. The asynchronous use of public key cryptography for signing large data volumes facilitates common attacks to breach signer's keys.   
\end{description}  

All of these threats bear the potential to seriously degrade the ICN service and lead to insufficient or erroneous data dissemination.  Another major risk for the ICN infrastructure---and from a general perspective for the ICN concept---results from the power that an end user gains over an (open) ICN distribution backbone.

\subsection{End Users Affect Backbone States}

\textbf{Content suppliers}\qquad
Related work on ICN security has primarily focused on validating content
correctness and authenticity. Commonly, self-cer\-ti\-fying security
credentials are included in `secure names' that facilitate mechanisms for
verifying authors, origins, and content integrity
\cite{wn-snin-10,dgoa-snni-10,gkrss-nca-11,fmg-psisa-11}. Thus a receiver
can be sure to obtain the correct content and an intermediate cache can
validate the correctness of the security credentials, which finally
prevents DoS on the ICN system \cite{gsksr-icnsf-11}.  Nevertheless, having
created (or learned) a valid name, any ICN member can re-announce this in
the request routing system or re-register it in the content resolution
service, thereby injecting poisonous routes or artificial names into the
system.\footnote{As a countermeasure, DONA introduces certificates of
publishers on the price of per cache-instance varying names. Content
routing then works on wildcarding names, which re-introduces the threat of
route poisoning.} Similar vulnerabilities of DNS and BGP are known from
today's Internet infrastructure \cite{bfmr-sbsis-10}, but remain restricted
to (topology) \emph{providers}. ICN opens the liberty of route injection to
the group of content suppliers (i.e., \emph{end users}). We will discuss threats unique to ICN in Section \ref{sec:route-map}.

Security aspects of Internet caching have been studied extensively, see \cite{dgck-padic-08} and refs.~therein. Current work has identified conceptual difficulties in preventing locality attacks, as they are hard to distinguish from regular user behaviour. It is worth noting that cache snooping raises privacy issues in ICNs, as well.  We will not dive deeper into this well-known subject. 

\textbf{Content consumers}\qquad
Little attention has been given to the effects of state management in ICN. Arianfar \emph{et al.} \cite{ano-crdi-10} discuss design choices for an ICN router. They concentrate on the content cache and explicitly do not consider per-request states. Perino and Varvello \cite{pv-rcccn-11} have evaluated requirements for content routers that hold content information bases in Bloom filters and reverse paths in pending interest tables (PITs). Under the assumptions of {\em valid} content requests propagated on {\em homogeneous} network links with a maximum global RTT of 80~ms, average PIT sizes are identified in the order of 1~Gbit/s for current line speeds. FIB sizes and lookup complexity were shown to depend nonlinearly on prefix numbers and name lengths. Lauinger \cite{l-sscn-11} explicitly addresses the threat of DoS attacks by filling the available memory of a router with pending interest states. 

Such attacks on hardware resources may be mitigated by limiting overall PIT sizes. However, securing router resources by table limits does not abandon resource exhaustion problems which equally occur whenever tables are filled. In the presence of a table limit, an attacker could initiate massive drops of PIs from a router's table and thus disrupt data delivery to regular receivers. The author in  \cite{l-sscn-11} proposes to drop interests at the head of the PIT, which however may easily be misused to DoS-attacking neighbors, or to use Bloom filters instead of PIs. If applied without strict capacity limits, the latter approach is vulnerable to flooding attacks as interface filters degrade their selectivity. 

Only recently, and after the first release of this technical report and its excerpt \cite{wsv-bipmc-12}, state management and related security issues have been discussed in  reports by Yi \emph{et al.} \cite{yamwz-csfp-12} and Gasti \emph{et al.} \cite{gtuz-ddnn-12}. The authors discuss core issues of route hijacking, state overload, and cache pollution and propose counter measures by extending interface functions, e.g., for limiting rates and survey content delivery. We will refer to those aspects in detail along the lines of this article.

\textbf{Intermediate Summary}\qquad
ICN opens the control plane of (ICN-enabled) backbone routers
for content consumers and suppliers (i.e., end users) on a fine-grained
base. This is a fundamental step away from the current Internet design and
bears significant risks. Current concerns in the context of
routing mainly focus on state explosion due to the large amount of content
items. One might argue that those resource exhaustions will be solved by
more powerful hardware in the future. We will discuss options and limitations of related core aspects in Section \ref{sec:basics}. Still, binding the integrity of
the routing infrastructure to end users activities is intrinsic to current ICN
approaches---and presumably to the overall ICN concept.

\section{Basic Threats to Stability}\label{sec:basics}

The control-plane in ICN will encounter changes and processing load for each publication and for each content request.  
Publication states require management in the request routing or mapping system, whenever a content item is published/withdrawn or a cache location is added/removed. Subscriptions initiate a grafting of distribution states throughout the forwarding system (DONA, NDN, NetInf)  or trigger optimized path finding procedures in the resolution service (PSIRP, NetInf).

In this section, we theoretically discuss resulting threats that inherently arise at the infrastructure level.

\subsection{Routing or Mapping Resources}\label{sec:route-map}

The common view on routing is that of a topological resolution service: Routing guides the paths to hosts. As ICN abandons the host-centric paradigm to address content objects directly, routes to content items attain the role of traditional topological directives.   

\subsubsection{State and Update Complexity}

In ICN, each content item (file) needs retrieval and therefore must be accessible via some resolution service. This may either be implemented by a distributed routing system, or by a mapping service that provides an indirection to topological locators of publishers or content caches.
The average complexity of the corresponding management operations reads $\langle\#~of~content~items\rangle~\cdot \linebreak \langle\#~cached~replica\rangle~\cdot \langle update~frequency\rangle$\ and must be considered a severe challenge. A global request routing system will need to host at least the amount of the Google index base (${\cal O} (10^{12})$) at a much enhanced update frequency (by caching). For comparison, today's DNS subsumes ${\cal O} (10^8)$\ names at a very low change rate of $\approx 10^5$ alterations per day. As a consequence, the request routing/mapping system is stressed by adding and updating name or cache entries at overwhelming frequency, the details of which depend on the implementation of the service. 
  
\subsubsection{Caching On-Path versus Off-Path}

Route maintenance in ICN consists of propagating content publishers (i.e., default paths) as well as  cache instances. While the first task is known to generate a high volume of data and frequent updates, caching is expected to largely exceed default announcements in number and update frequency. 
As a countermeasure, data replication  may be limited to caching along default paths, which remarkably reduces the complexity for the routing system. On-path cache replica are met implicitly when requests are routed towards the source, and need not be advertised in the routing or mapping service.  On the downside, restricting the caching to default paths will drastically reduce its effectiveness, and a corresponding strategy falls behind today's CDN solutions. Godsi et al. \cite{gsksr-icnsf-11} discussed the caching problems in detail. The authors came to the conclusion that on-path caching is merely a warm-up of traditional Web proxies.

\subsubsection{Route Integrity}

ICN, like the current Internet, relies on the integrity of its routing system. A bogus route may block or degrade services, lead to incorrect content delivery, or violate privacy. These core concerns are well-known from BGP \cite{bfmr-sbsis-10}. Without considering protective measures in BGP, Yi et al. \cite{yamwz-csfp-12} and Gasti et al. \cite{gtuz-ddnn-12} compare BGP with NDN security. The authors argue that the NDN approach reduces vulnerability to black-holing, as routers can identify unresolved content requests and rank/re-route per prefix and interface. 
Nevertheless, these countermeasures cannot prevent attacks of interception and redirection with service degradation. 

Aside from those vulnerabilities known from BGP routing, the following threats uniquely arise for content-centric routing.

\begin{description}
\item[Route Set Inflation] Almost all ICN solutions allow for universal caching without explicit authorization. Any in-network cache can thus announce any content name, while origin validation measures such as RPKI \cite{draft-ietf-sidr-pfx-validate} cannot be applied. 
\item[Route-to-Death] Route redirections may be applied to slow down content delivery or to jitter response times. As the routing infrastructure is vulnerable to increased delays and delay variations, corresponding threats arise (see Section~\ref{sec:forwarding-threats}).
\end{description} 

\subsection{Forwarding Resources}
\label{sec:forwarding-threats}

Traditional routers in today's Internet consist of a central processing unit and main memory that are available to the control plane, mainly to learn and determine new routes, as well as FIB memory that is fed by the route selection process. Data forwarding remains bound to  FIB lookup and packet processing at line-cards. This design choice purposefully decouples forwarding capacities from control processing and--with equal importance---protects control states from (bogus) data packets.

Current concepts of content-centric data forwarding break with this separation paradigm, and introduce---similar to multicast---an additional reverse path forwarding table, also called PIT. Unlike in multicast, this table is updated {\em packet-wise} on line speed by data-driven events. In the following subsections, we concentrate on the consequences for routing resources  in detail. We will consider a  chain of routers $P_i$ along a data path and use the notation summarized in Table \ref{tab:notation}.

\begin{table}
\centering
\begin{tabular}{|c|p{0.8\columnwidth}|}
\hline
$R_i$ & The $i$-th Router\\
$C_i$ & Capacity of the link between $R_i$  and $R_{i+1}$\\
$U_i$ & Utilization of the link between $R_i$  and $R_{i+1}$\\
$S_i$ & \# of content request states of $R_i$ at its interface towards  $R_{i+1}$\\  
$\alpha_i$ & Content request rate at interface $R_i \rightarrow  R_{i+1}$\\
$\omega_i$ & Content arrival rate at interface $R_i \leftarrow  R_{i+1}$ \\ 
$T_i$ & Request timeout at interface $R_i \rightarrow  R_{i+1}$ \\
$l$ & Packet length \\
$\langle \cdot \rangle$ & Average value of $\cdot $ \\
$\sigma(\cdot)$ & Standard deviation of  $\cdot $ \\
\hline
\end{tabular}
\caption{Glossary of Notations}\label{tab:notation}
\end{table}

\subsubsection{Content Request States versus Content Request Rates versus Network Utilization}\label{sec:state-rate}

Content request states, i.e., Pending Interests in NDN, are the essential building block to flow control in a content-centric distribution system that operates hop-by-hop. Each request state will trigger a data packet on return, why the number of open request states corresponds to data arrival at this interface after the transmission time.

Consider a point-to-point interface between routers in steady operation and in the presence of a (per interface) state timeout $T_i$. In the absence of request retransmissions, packet loss, and state dismissal, we first want to derive the relation between router states at time $t$  and network capacity. The basic rate equation reads

\begin{eqnarray}
S_i (t) & = & S_i (t - T_i) + \int_{t-T_i}^t \alpha_i(\tau) - \omega_i(\tau) d\tau\nonumber\\
 & = & S_i (t - T_i) + \int_{t-T_i}^t \alpha_i(\tau) - \alpha_i(\pi(\tau)) d\tau\nonumber\\
 & = & \langle \alpha_i \rangle \cdot \min(\langle RTT\rangle, T_i) + \nonumber\\
  & & {\cal O} \left( \sigma( \alpha_i) \cdot \sigma(\min( RTT, T_i))\right),
\label{eq:rate}
\end{eqnarray}

\noindent where  $\pi(.) $ denotes the time shift corresponding to the packet arrival process and  $RTT$ the random variable of packet round trip times, which is assumed independent of the requests and packet rates. 


From Equation (\ref{eq:rate}), we can immediately deduce that timeout values below the (varying) $RTTs$  limit the number of request states, but at the same time will block data forwarding. 
A second view reveals the strong dependence of states on the $RTT$ variation. A similar phenomenon is well-known from TCP \cite{j-cac-88}, but has been overlooked in corresponding previous work on ICN resource considerations \cite{pv-rcccn-11,yamwz-csfp-12,gtuz-ddnn-12}. 
 
Henceforth we will address the case of data flowing unhindered by the state timeout $T_i$. Furthermore---for a steady-state scenario---it is assumed that the content request rate fluctuates on the same stationary scale. Equation  (\ref{eq:rate}) then simplifies to 

\begin{eqnarray}
S_i (t) & \approx &  \langle \alpha_i \rangle \cdot \left(\langle  RTT\rangle + \kappa\,  \sigma( RTT) \right)\label{eq:state-estimator}\\
 & \approx & U_i(t)/{\langle l \rangle} \left(\langle  RTT\rangle + \kappa\,  \sigma( RTT) \right),
\label{eq:state-util}
\end{eqnarray}

\noindent with an estimating parameter $\kappa$ for the mean deviation.\footnote{The corresponding (over-)estimator in TCP is commonly set to $4$. However, it is well known that standard TCP algorithms and parameters are inefficient at rapidly changing round trip times, which are characteristic for interface conditions in content-centric routing.} For the last step, we roughly assumed that content requests and content arrival are in stationary equilibrium. 

Approximation (\ref{eq:state-util}) yields the desired coupling of the link utilization and the state management resources at a router: On a single point-to-point link without state retransmissions and in flow balance, state requirements are proportional to the  network utilization, enhanced by a factor of a {\em global retransmission timeout.} At switched interconnects or in bursty communication scenarios, conditions are expected to grow much worse.

The following observations are noteworthy.
\begin{enumerate}
\item Unlike in TCP that estimates a single end-to-end connection, content request states at routers aggregate prefixes and numerous flows to various content prefixes. Moreover, content items (prefixes) are explicitly not bound to end points. Thus rapidly varying RTTs are characteristic to interfaces and even to flows in content-centric routing. The presence of  chunk caching  may even increase the $RTT$ variation. Hence, no convergent estimator for a round trip time can be reasonably given.\footnote{Yi et al. \cite{yamwz-csfp-12} propose to maintain RTT estimators per prefix and interface in the FIB, which would find its analogy in BGP by adding RTT estimators per IP prefix. The evaluation of such quantities would not only require an extensive pairwise probing, but at the end is not well defined for (aggregated) prefixes.} 
\item In the current Internet, the variation of $RTT$ is commonly larger than its average. End-to-end delays are known to approximately follow a heavy-tailed Gamma distribution \cite{bmhum-aeedm-02}. PingER \cite{pinger} reports means and standard deviations of about 250~ms, with maxima up to 5,000~ms. 
\item Limiting the absolute size of the content request table imposes a strict bound on network utilization. However, the  sustained rates are mainly determined by actual RTTs and are hardly predictable.
Similar arguments hold for defining timeout values.
\item A corresponding picture arises, when content request rates are limited instead of table sizes.\footnote{Rate limiting has been currently proposed in \cite{yamwz-csfp-12}.} For an 'on average' optimal limit $C_i \cdot \langle RTT \rangle/\langle l \rangle$, the variation of content replies in time may lead to large over- and under-utilization of network resources that goes along with large fluctuations in request table sizes.
\end{enumerate}

\subsubsection{Memory Requirements}

A content-centric router that is designed to fully utilize its link capacities, requires sufficient table space for content requests under varying network conditions. Equation (\ref{eq:state-util}) approximates the corresponding resources when applied to the maximum link capacity $C_i$. Using the conservative value of $\kappa = 4$ as for TCP, a packet length $l =$ 1,000~bytes,  and $RTT$ values from PingER as cited in the previous section, we derive

\begin{equation}
S_i = 1,25\, s / 8.000\, bit \cdot C_i \approx 1,6 \cdot 10^{-4} {s}/{bit}\cdot C_i
\label{eq:memory}
\end{equation}

For a line-speed of 1 to 100 Gbit/s, 160k  to 16,000k content request entries then need to be installed per interface at minimum.\footnote{This number  may well underestimate the need by an order of magnitude or two.} Due to the more accurate inclusion of $RTT$ variation terms, these findings differ from previous results \cite{pv-rcccn-11,yamwz-csfp-12} by more than an order of magnitude. Still they are merely a rough estimate, as larger fluctuations of round trip times may significantly increase resource demands.

It is noteworthy that Equation (\ref{eq:memory}) holds for any router in a content-centric Internet. Unlike today, where full BGP tables are only required at AS border routers, and interior devices operate on a very small routing table, CCN access routers already demand for a full PIT memory, the size of which is determined by its interface capacities. In practice, this significantly increases router costs, as any fast interface must co-locate a large block of fast memory.

\subsubsection{CPU Load from Table Management}

An ICN content router maintains states according to user data requests. For any content request, it needs to (1) insert a state in its request table. On the arrival of any data packet, it needs to (2) search and (3) delete on success in the same table. These operations are required at line speed. In addition, a router has to (4) maintain timers of all (soft) states in its content request table. 

The paradigmatic shift of ICN opens the state table, which is responsible for data delivery, to the end users. Any content requester may create unexpected access patterns. Consequently, to guarantee robustness, an implementation of the request table not only needs to perform dictionary operations very efficiently {\em on average} but also in worst-case.  As there is no final design of an ICN router, we discuss the general design space for hash tables in this context.

The most efficient implementation is the usage of on-chip content-addressable memory (CAM), which is the complete hardware counterpart of a dictionary data structure. However, costs, energy, and limited size prohibit its deployment for expected PIT tables sizes.
Typically, state tables for request routing (or caching) are implemented using hash tables. In its elementary form, these data structures can perform all basic dictionary operations (i.e., insert, search, and delete) at constant complexity, but experience hash collisions. Collisions cause a conflict: An implementation that ignores hash collisions will overwrite data. This limits the field of application for ICN, as dealing with collisions increases complexity, or making the system directly threatened by DoS.

In ICN, Perino and Varvello~\cite{pv-rcccn-11}, for example, propose to use HC-basic~\cite{bppp-hcsnb-09}, a collision-prone scheme without avoidance mechanisms.  Such a design choice makes the ICN system vulnerable to (also purposefully) replacing valid content requests. In realistic deployments of ICN, an implementation of hash tables will either provide mechanisms to prevent or to handle hash collisions. 

Essentially four solutions are known to overcome hash collisions: (1) Hash chaining or open addressing, (2) perfect hashing, (3) cryptographic hash functions, and (4) universal hashing.

Hash chaining (i.e., concatenating conflicting keys) or open addressing (i.e., deterministic probing for an alternate location) \cite{clrs-ia-01} circumvent collisions on the price of enhanced update costs. The worst case complexity increases to ${\cal O}(N)$ for a table of size $N$. This introduces  well-known vulnerabilities, as any pattern that creates collisions will result in such linear complexity instead of amortized ${\cal O}(1)$. Crosby and Wallach~\cite{cw-dsaca-03} analyzed this for current software systems (e.g., the Bro IDS). For the more sophisticated and widely deployed hardware hash tables Peacock and Cuckoo, Ben-Porat \emph{et al.}~\cite{bblp-vhhts-12} recently studied the structural vulnerability and observed significant performance degradation, as well. Applying these approaches to ICN introduces an obvious threat.


``Perfect hashing'' \cite{clrs-ia-01} supports constant complexity in the worst-case, but requires a static key set. It is thus not suitable for dynamic content requests that are characteristic in ICN.

Collision resistant cryptographic hash functions can be applied, but lead to a prohibitive increase in memory and CPU consumptions. Enabling cryptographic operations at backbone routers has been discussed continuously in the context of securing BGP~\cite{bfmr-sbsis-10} and did not succeeded so far. In contrast to requests in ICN, BGP updates occur rarely (even if we consider update storms) and will be sent by known peers. ICN would have to apply cryptographic hashing to all Interest packets. Cryptographic hash functions are---with respect to the packet processing requirements in ICN and current, long-term hardware development---out--of--scope also for a future implementation of the ICN paradigm.

A trade-off between performance and worst-case avoidance can be implemented by universal hashing \cite{cw-uchf-79}. It has been introduced to obfuscate the hash function by randomly selecting the hashing algorithm at start-up. Unfortunately, universal hashing is difficult to implement in hardware. Moreover, it still shows a collision probability which is small compared to basic hashing but high compared to cryptographic hash functions. Threats based on fingerprinting make the system additionally vulnerable. It should also be noted that universal hash tables cannot be deployed in a distributed fashion, but are confined to strictly local use cases due to the random selection of hash function. Collaborating request management is thus challenged.


For a detailed overview on state-of-the-art hash tables for high-speed packet processing, we refer to the excellent work by Kirsch \emph{et al.} \cite{kmv-hthpp-10}. Even though wire-speed hash tables are deployed in the current Internet, it is by no means obvious that these approaches provide sufficient robustness in daily ICN operation. Current solutions exhibit serious attack vectors, either by DoS collision overwrite or by a non-constant worst-case performance. In addition, even a constant complexity does not guarantee appropriate robustness as extra memory accesses can prevent wire speed performance.

\subsection{Discussion}

ICN approaches introduce three major changes to the current, well approved Internet infrastructure:

\begin{enumerate}
\item Information objects become first class routable network entities.
\item The control plane of the routing system is opened to the end users, who are entitled to implement states at routers that are managed by data-driven events. 
\item The end-to-end design principle is abandoned in favor of of a hop-by-hop routing coordination. 
\end{enumerate}

All of these novelties impose severe challenges to an infrastructure serving the needs of a future content-centric Internet. While the first change threatens an infrastructure design by the mere amount of  entities to be accommodated, an opening of the control plane gives rise to manifold attack vectors. In such an environment of a  open access to infrastructure states, security and robustness measures need to be applied to many more constituents of the distribution system than in today's Internet. In addition, the requirement of managing states in line with data forces any of the (user-centred) operations at the control plane to run at wire speed. The latter is an exceptionally hard challenge, as even operations of constant complexity may exceed the realm of wire speed 
processing. 

Finally, the hop-by-hop approach applied in contrast to the end-to-end concept promises the advantage of early adaptation to changing network conditions. However, the problem of resource coordination makes hop-by-hop forwarding significantly less robust than solutions built upon the end-to-end design principle. In fact, it is very easy to construct cross-traffic scenarios that lead to disastrous performance of hop-by-hop forwarding paths. Lead by this insight, Carofiglio et al. \cite{cgm-jhric-12} propose to build ICN on a hybrid solution of AIMD (additive increase, multiplicative decrease) applied to intermediate routers {\em and} end points. However, the authors did not account for effects of delay and delay variation between content requests and content fulfillment. In addition, aggregation needs to be taken into account that spoils the concept of adapting to individual flow parameters.

When examined from the conceptional side, ICN solutions raise many fundamental problems and---up until now---fail to resolve the corresponding problems. It is the aim of the present paper to identify challenges  and sharpen the view on issues that require resolution. Detailed solutions of the identified problems are out of scope for the present work.

In the following, we will experimentally evaluate the content delivery system for the example of name-based routing. To analyze the behavior of the corresponding infrastructure, we perform
only common ICN operations (i.e., subscriptions of existing and
non-existing data) in a straightforward topology with a simple content
model. This allows us to omit side effects due to complex routing or
caching issues.

\section{Experimental Evaluation}\label{sec:evaluation}

In this section, we present the results of straight-forward experiments that show the outcome of the core threats as theoretically discussed in Section \ref{sec:basics}. In particular, we concentrate on system and performance implications of the data-driven state management at infrastructure devices.  Even though the measurements mainly relate  to the NDN implementation \texttt{ccnd}, we should emphasize that we do not evaluate the implementation itself, but use it as one
real-world instance of the information-centric network deployment to illustrate the routing protocol mechanisms. Following this spirit, we do not interpret or discuss absolute performance values, which surely can be improved by  optimized software and hardware in the future, but focus on structural and asymptotic analysis.

\subsection{Core Measurement Setup}\label{sec:setup}

\begin{figure}
  \includegraphics[width=\columnwidth]{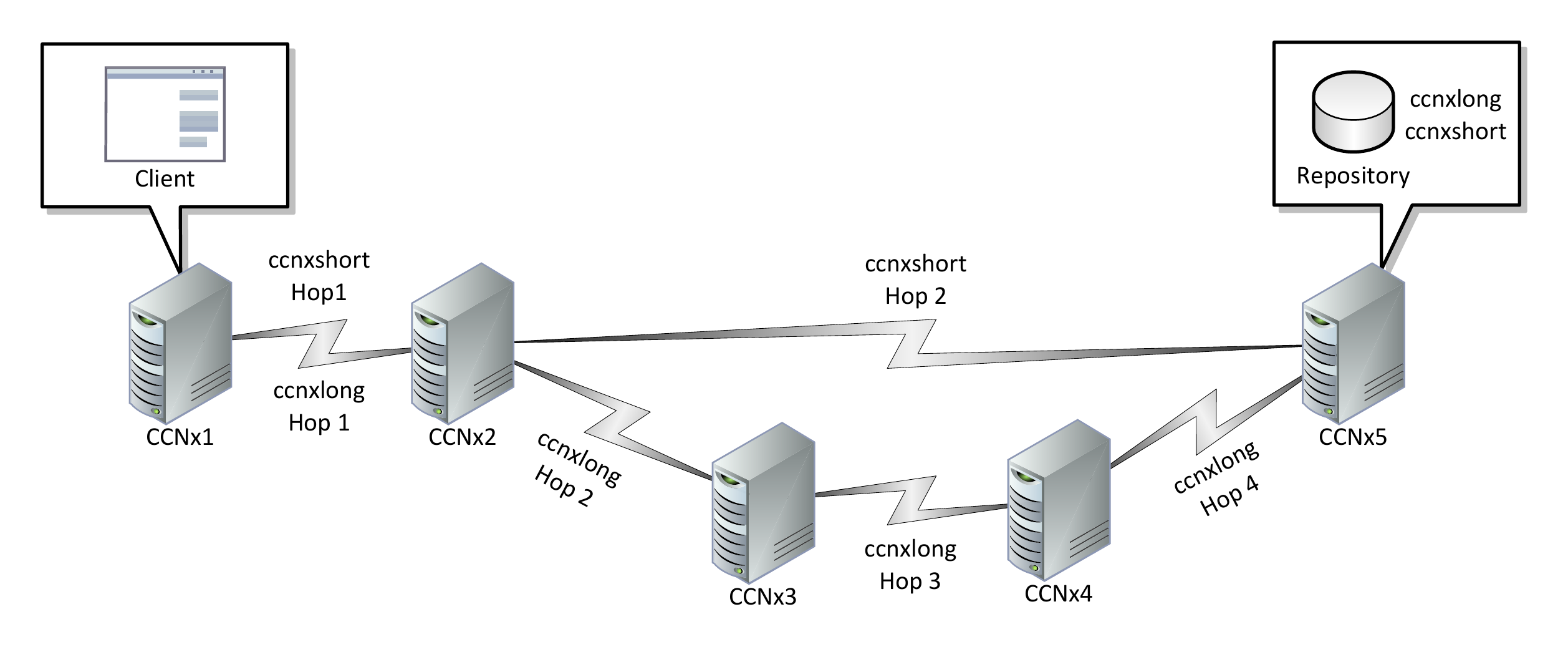}
  \caption{Topology in the experimental setting}\label{fig:topology}
\end{figure}

In our measurement study, we purposefully deploy {\em simple} communication scenarios between one content requester and one publisher. The basic network topology is represented by a Daisy chain of directly
interlinked CCNx routers with 100 Mbit/s, one end connects the content
consumer and the other the content repository (see Fig.~\ref{fig:topology}). The \emph{basic topology}
consists of two hops and the \emph{extended topology} of five nodes. It is noteworthy that more complex settings, e.g., a Dumbbell topology popular to represent backbone network effects, would enforce the effects, which we already see in our simpler and more transparent examples.

We use the CCNx implementation version
0.5.1~\cite{ccnx}, i.e., the client library to announce content Interests,
the content repository to store data, and the \texttt{ccnd} to forward
subscription and data. The following analysis focuses on the effects
on the router side. For obtaining a fine-grained view, we concentrate
on the local system as well as inter-router dependencies.

For all CCNx parameters, we leave default values. In particular, routers do not follow a specific strategy layer, as this would push resource consumption towards specifically twisted but equivalent results as discussed in Section~\ref{sec:ps}. This also implies deactivated rate limiting, which additionally introduces further attack vectors (cf., Section~\ref{sec:attackinfiltration}). CCNx routers communicate via TCP (preserving packet order in the basic experiments) or UDP (extended experiments).

\subsection{Resource Consumption: Basic  Experiments}\label{sec:eval-simple}
\subsubsection{A Fast Path to Resource Exhaustion}\label{sec:resourceex}

An elementary threat intrinsic to data-driven state management arises from the overloading of routers by Interest
requests. This is most easily provoked by initiating requests for content that does \emph{not} exist. In our scenario, the consumer
issues 2,000~Interest messages for \emph{non}-existing content, waits
6~seconds, and repeats these steps until overall 150,000~Interests have
been sent.

\begin{figure}
  \includegraphics[width=\columnwidth]{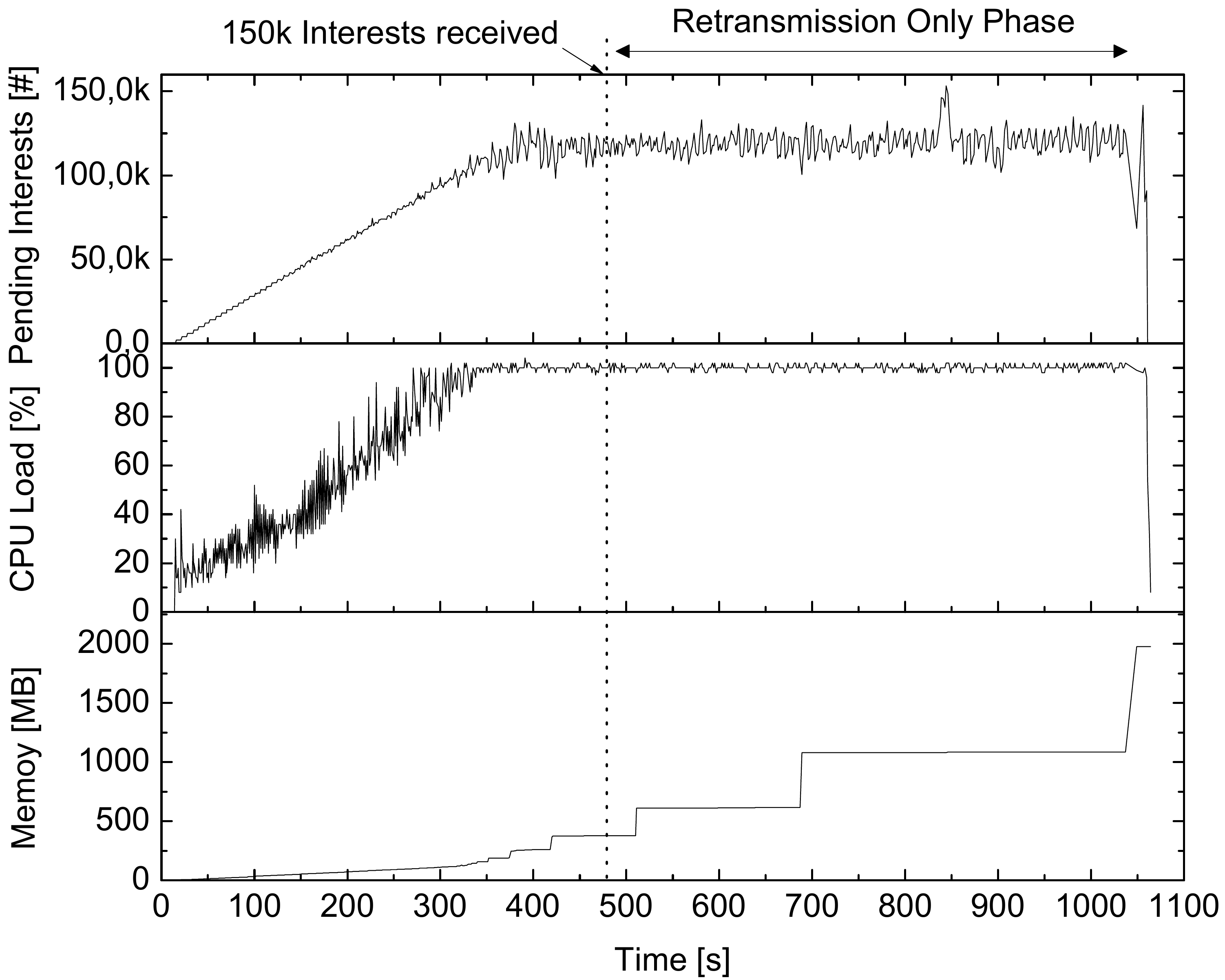}
  \caption{Load at the designated router of the receiver while requesting non-existing content}\label{fig:nocontent-150000interests}
\end{figure}

Figure~\ref{fig:nocontent-150000interests} shows the local resource
consumptions on the first hop of the content receiver. The number of
entries in the Pending Interest Table (PIT), the CPU load, and the required
memory increase linearly with subsequent bulks of Interest messages until
the system is saturated. In this case, the router reaches its limits of 
processing and memory resources when storing $\approx$\,120,000~PIT
entries. While sending Interests, the initiating node retransmits previous
announcements to keep states fresh at the router. Even though the
retransmission timer is below the expiration timer and network delays are
very short, the PIT size fluctuates as entries drop due to overloading.
After all initial Interest messages have been distributed, the content
consumer only retransmits subscriptions.

Our experiment illustrates several problems: A router may easily exhaust PIT space, but even if it was able to store
all entries, it would suffer from a `retransmission only' phase. The
retransmissions agglomerate over time and create a continuous stream of
signaling that consumes CPU cycles. When the update rate is higher
than the processing capabilities permit,  retransmissions require 
buffering, which leads to additional memory overhead (cf.,
Figure~\ref{fig:nocontent-150000interests}). A high system load increases
the probability of dropping a PIT entry even if its refresh message has 
been signaled in time. This again causes additional operations on the PIT
data structure (add/ delete calls) and fosters the momentum of load.

\begin{figure*}[ht]
  \subfigure[2 files per second]{\includegraphics[width=0.33\textwidth]{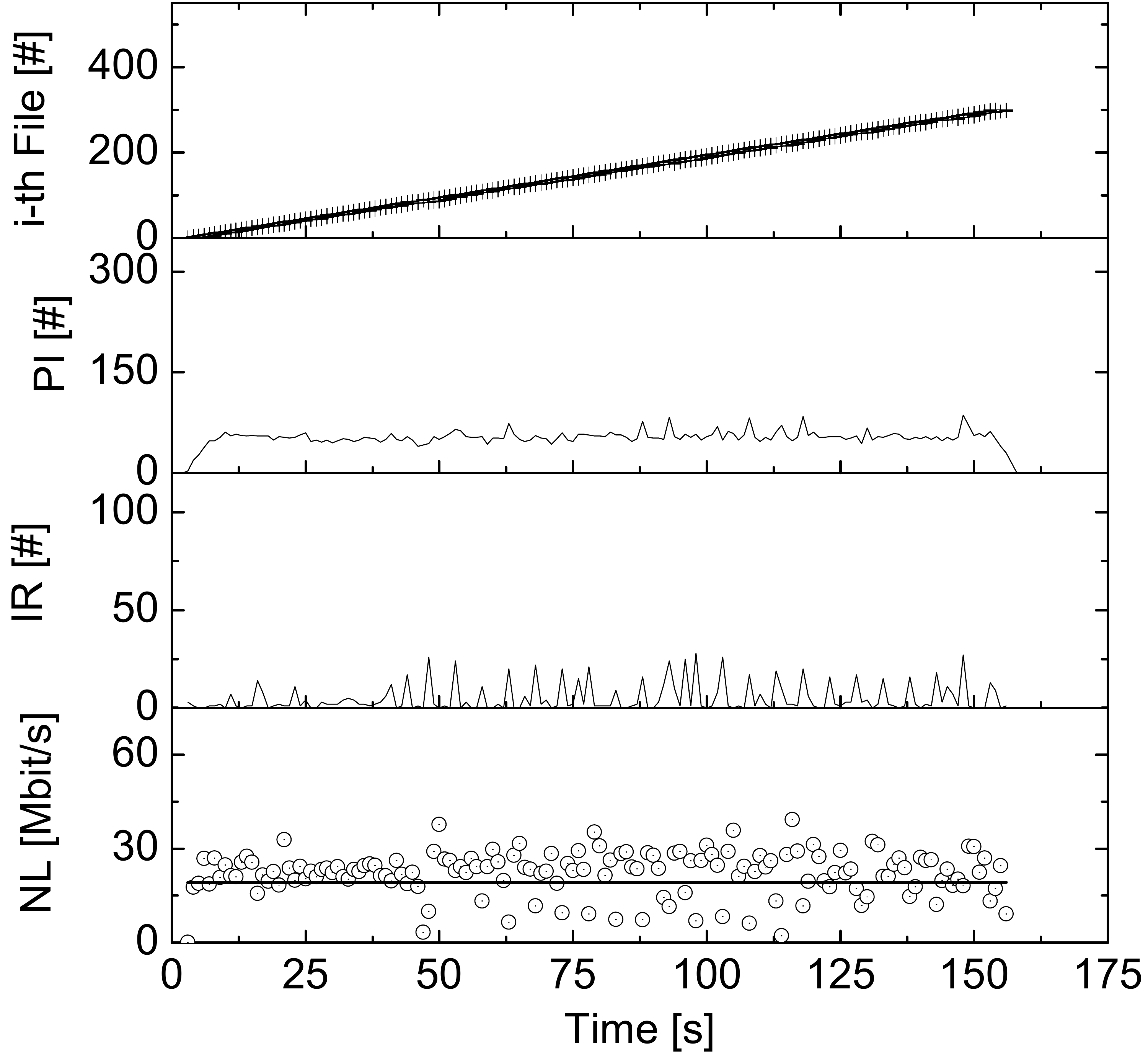}\label{fig:parallel-download:2files}}
  \subfigure[10 files per second]{\includegraphics[width=0.33\textwidth]{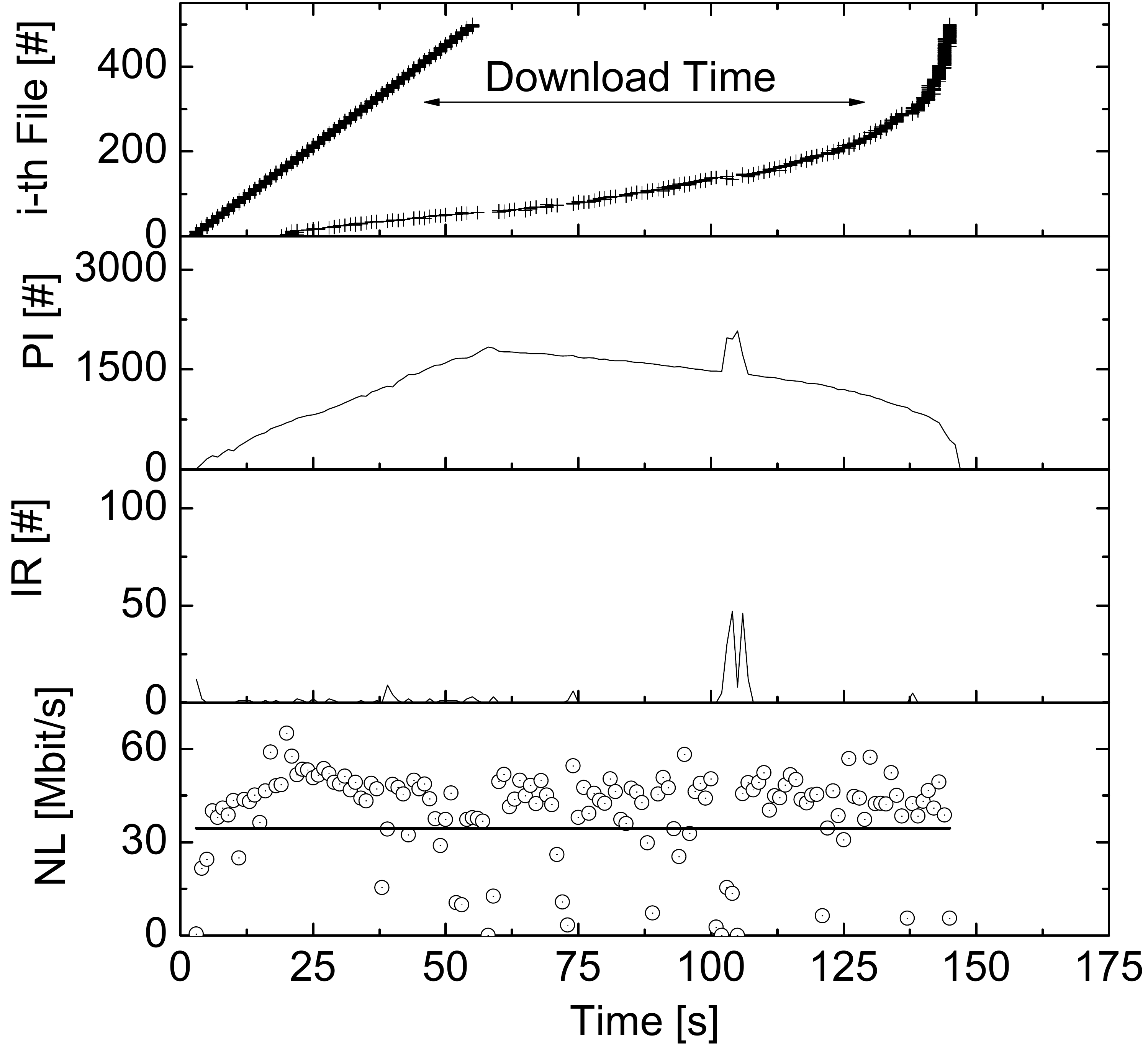}\label{fig:parallel-download:10files}}
  \subfigure[100 files per second]{\includegraphics[width=0.33\textwidth]{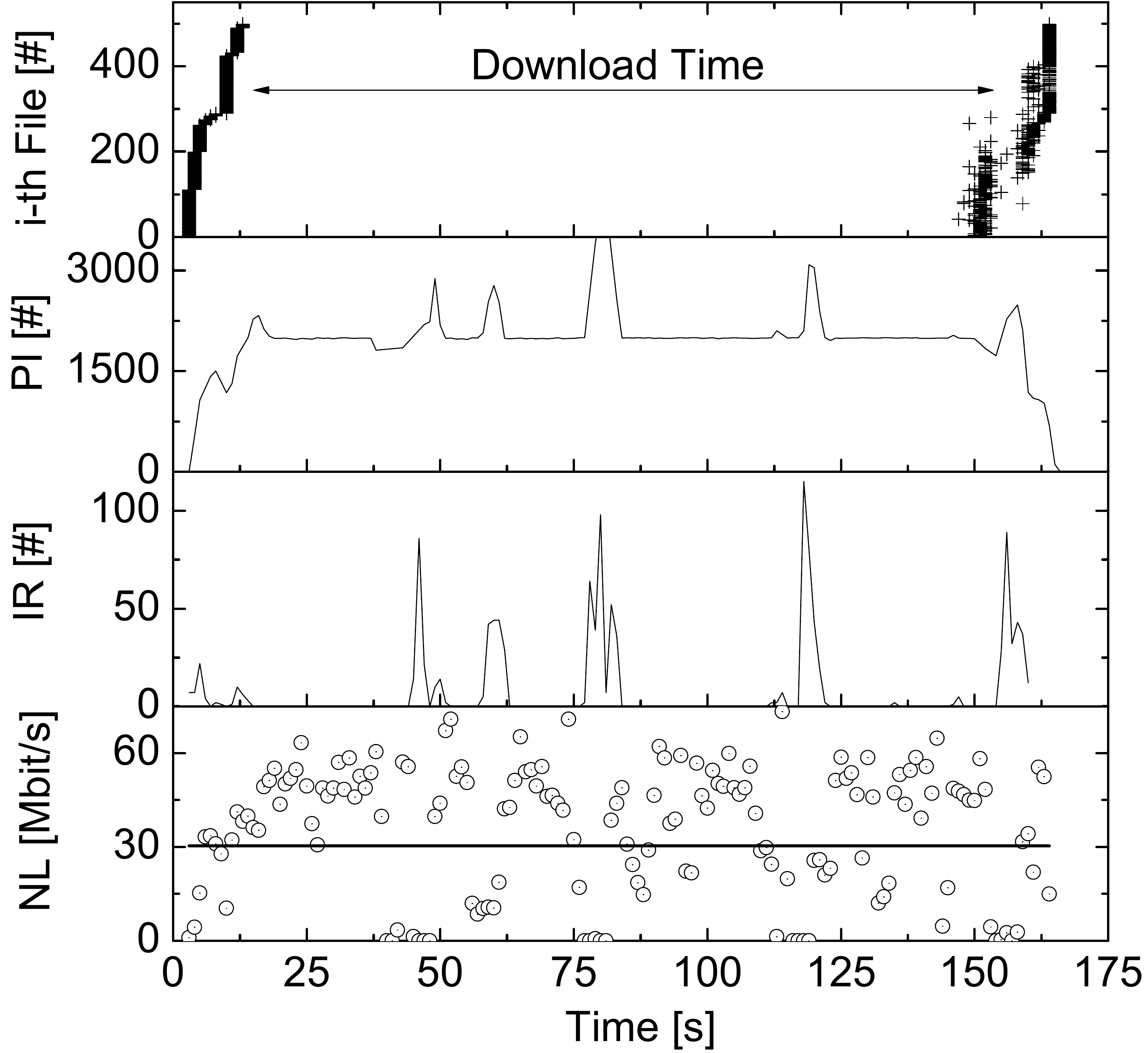}\label{fig:parallel-download:100files}}
  \caption{Parallel download of 10~Mbit files: Start and stop time of the
  download per file at the receiver \& resource consumption at its
  designated router [Pending Interests (PI), Interest Retransmits (IR), and
  Network Load (NL) including the mean
  goodput]}\label{fig:parallel-download}
\end{figure*}

In a recent publication, Yi \emph{et al.} \cite{yawzz-afndn-12} propose to mitigate this threat by signaling content unavailability back to the original requester. Such  \texttt{NACK} will cure the Interest retransmission effects discussed above, while at the same time \texttt{NACK} suppression introduces a new attack vector at the supplier side. Unhindered by the presence of \texttt{NACK}, a bogus requester still can harm the routing infrastructure (in particular its designated router) by subsequent new Interest messages for unavailable content. 

\subsubsection{Chunk-based State Multiplication}


To analyze the performance of content consumption, we conduct a bulk file
transfer. At this, the content receiver initiates the parallel download of
multiple 10~Mbit files over a constant time. We consider three extremes, the
request of 2~files, 10~files, and 100~files per second.
Figure~\ref{fig:parallel-download} shows the start and completion time of the
download per file (top graph), as well as the Pending Interest Table (PIT)
size, the effective number of Interest retransmissions, and the traffic load
including the mean goodput at the first hop. For visibility reasons, we
rescaled the y-axis of PI in Figure~\ref{fig:parallel-download:2files}.

With an increasing number of parallel downloads not only the download time
increases significantly, but also the interval of the request and receive
phase grows in the scenarios of overload. While the download time is almost
constant for two files per second (cf.,
Fig.~\ref{fig:parallel-download:2files}), the stop time diverges
non-linearly from the beginning of the download in the cases of excessive
parallelism (cf.,
Fig.~\ref{fig:parallel-download:10files},\subref{fig:parallel-download:100files}).
150~s are needed to download {\em each} single file in the worst case
(Fig.~\ref{fig:parallel-download:100files}), while the link capacity would
permit to retrieve {\em all} files in about 10~s.

The reason for this performance flaw is visualized in the subjacent graphs.
A higher download frequency leads to an increasing number of simultaneous
PIT entries, which require coordination with the data plane. Each file
request will be split into requests of multiple chunks, in which the
generation of corresponding Interest messages will be pipelined. In
contrast to Section~\ref{sec:resourceex}, content exists. As soon as the
content traverses, Interest states dissolve and thus release memory.  These
operations cause a simultaneous burst in CPU load (not shown) and result
in growing Interest retransmits after droppings or timeouts (shown in
second lowest graphs). This also leads to retransmissions of data chunks.
As an overall net effect, the network utilization fluctuates significantly,
but does not adapt to actual user demands: Even though data requests could
fill the links easily, the average load remains about constant at 30~\% of the total network capacity.

In this example we demonstrated that insufficient processing and memory resources will strictly prevent a proper link utilization. This problem cannot be mitigated by rate limiting, as reduced Interest transmission rates will simultaneously reduce network utilization even further (see Section \ref{sec:state-rate}).\footnote{We should remind that applying Interest rates in NDN is a mechanism of flow control, and {\em not} for system resource protection. Intermingling these two aspects is likely to produce unwanted performance flaws and leads to new attacks (cf., Section~\ref{sec:attackinfiltration}).} 

The only visible way to assure proper  utilization of network resources requires appropriate routing resources, i.e., a PI table implementation that is sufficiently large and reliably operates at line speed. As we learned from the analysis in Section \ref{sec:forwarding-threats}, corresponding solutions are not available today. At the current state of the art, an attacker  can always reproduce the performance degradations by either blowing up RTT and its variation, or by injecting states that degrade the performance of the PI hash table of the routers.










\subsection{State Propagation and Correlation: \\ Extended Experiments}\label{sec:eval-extended}

\begin{figure*}[ht]
  \subfigure[Pending Interests]{\includegraphics[width=0.33\textwidth]{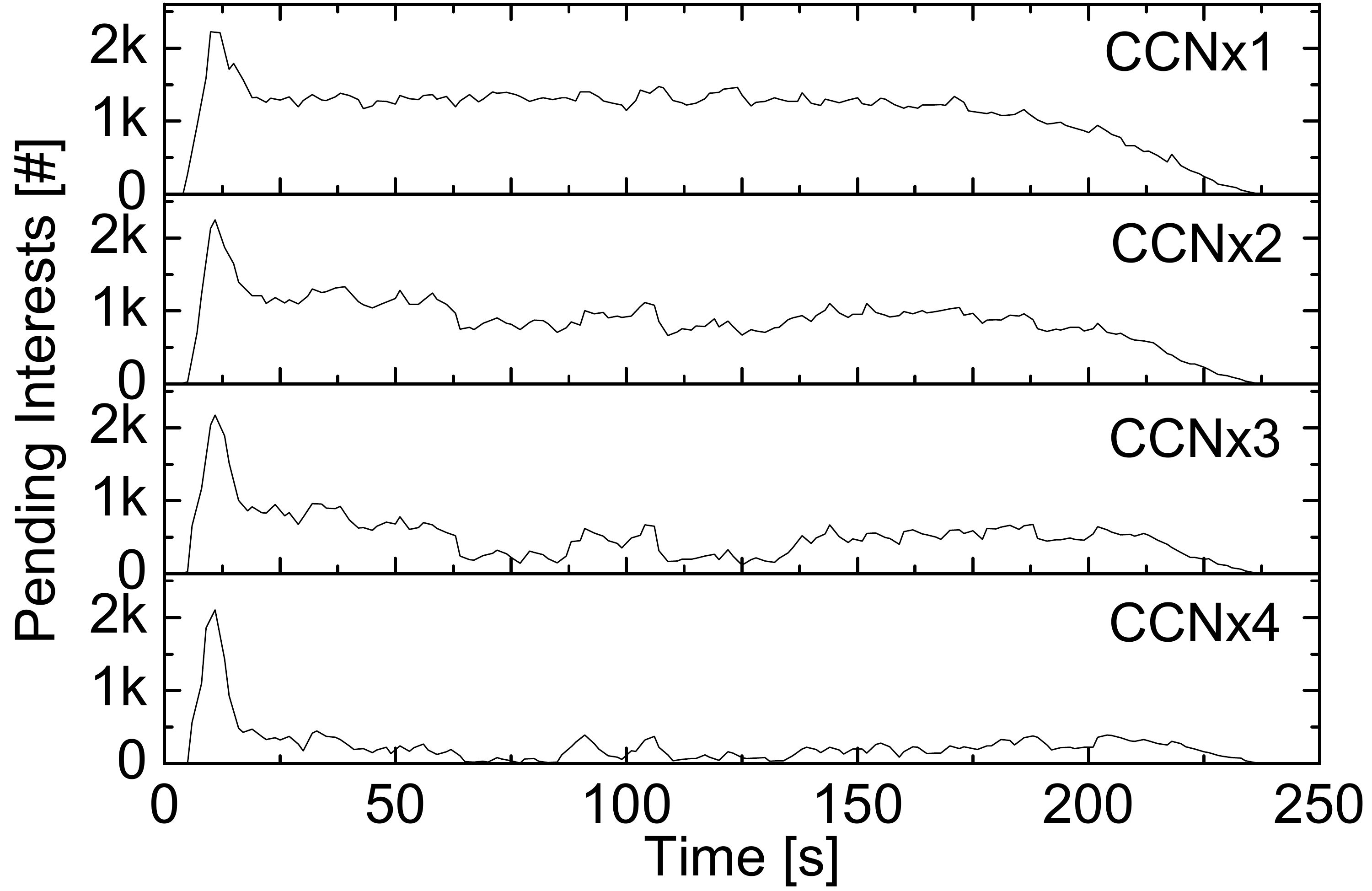}\label{fig:hom_chain_pi}}
  \subfigure[Interest Retransmits]{\includegraphics[width=0.33\textwidth]{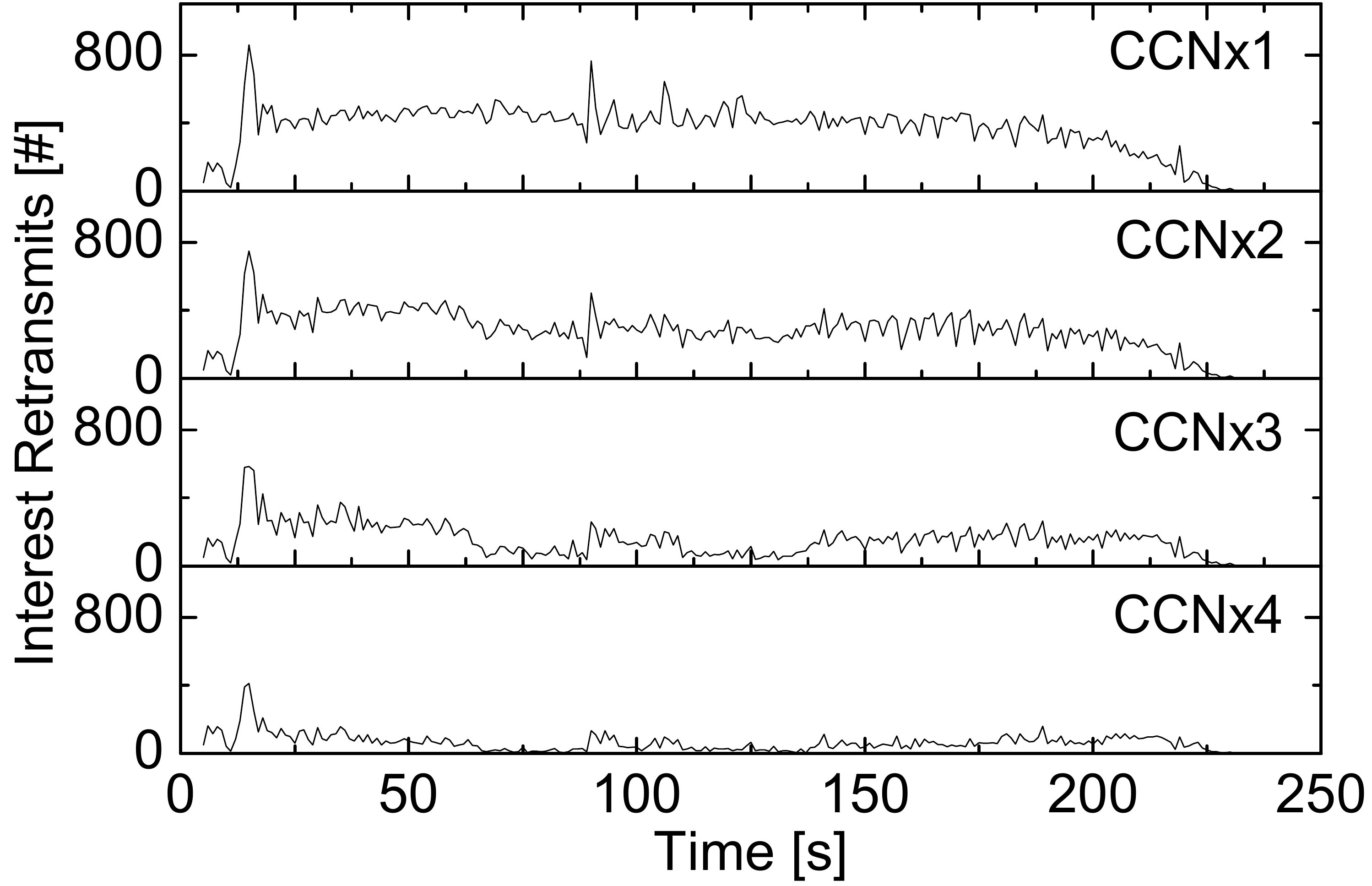}\label{fig:hom_chain_re}}
  \subfigure[Network Utilization]{\includegraphics[width=0.33\textwidth]{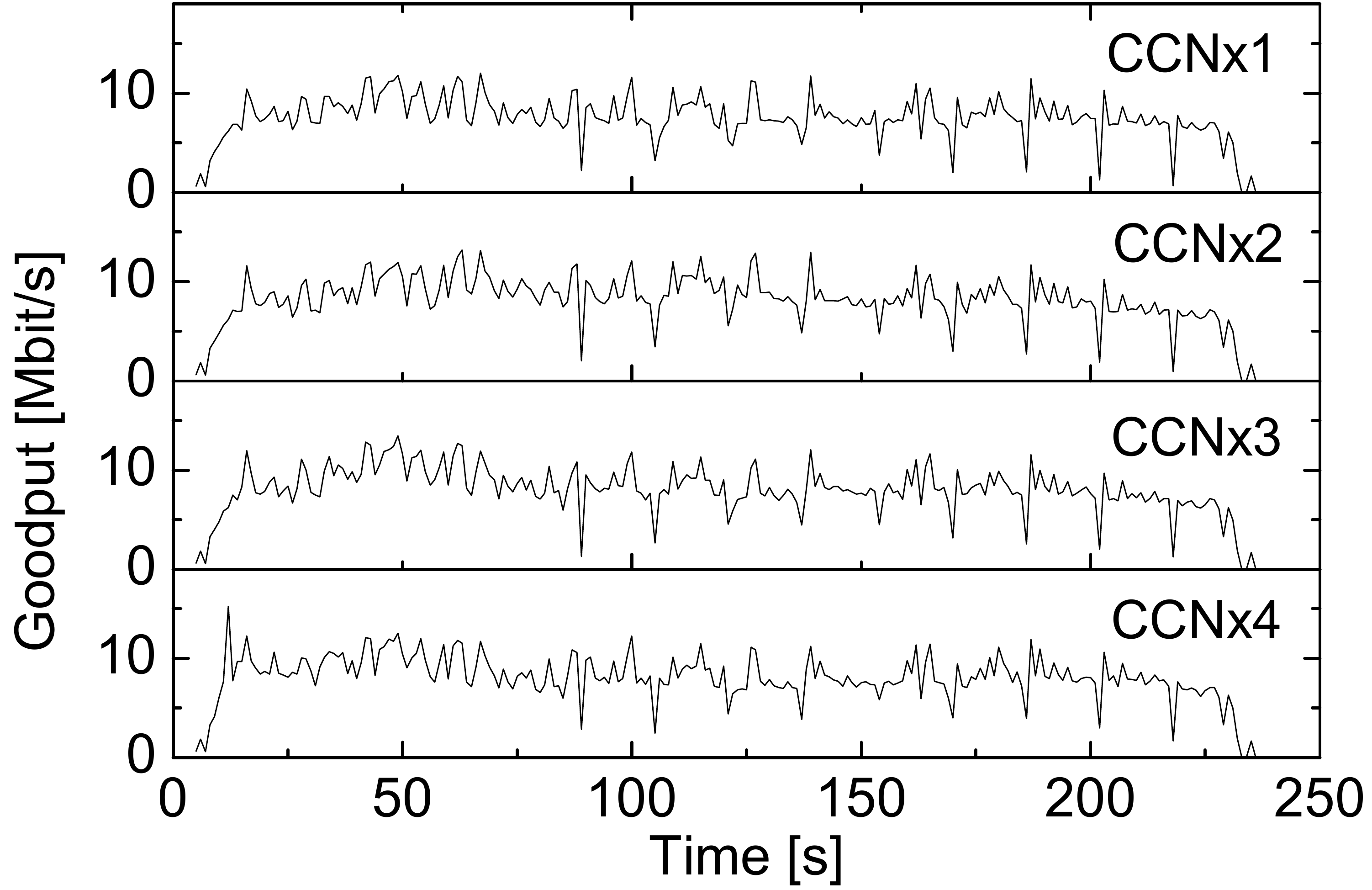}\label{fig:hom_chain_gp}}
\caption{Routing and forwarding performance in a homogeneous five-hop network} \label{fig:hom_chain}
\end{figure*}

In our extended experiments, we take a closer look at hop-by-hop routing performance using the five node routing chain displayed in the lower part of Figure \ref{fig:topology}. Intermediate nodes are numbered from the designated router of the content receiver
(first hop) to the router of the content repository (fifth hop). In the first step, we specicically concentrate on correlation effects of a single requester/publisher pair, thereby neglecting any competing cross traffic. Instead, we steer and survey routing resources in our experimental setup by using parametrizable virtual machines.

\subsubsection{Extending the Easy: A Homogeneous Network}

In this first extended experiment, we simply move our previous picture to the larger topology. All forwarding nodes offer the same resources, two cores@2.4~GHz, 3~GB RAM, and link capacities of 100~Mbit/s. A content requester downloads 500 files of size 10~Mbit at an average rate of 100~files per second. 

The corresponding results are visualized in Figure~\ref{fig:hom_chain}. It is nicely visible, how Interest state and retransmission management propagate through the network, showing an overall flattening towards the source, because states resolve earlier from faster packet delivery. Resource challenges due to state management as discussed in Section~\ref{sec:eval-simple} lead to an accumulated reduction in network utilization. The network goodput decreases from $\approx$ 30\,Mbit/s to $\approx$ 8\,Mbit/s, a reduction that is about linear in the number of hops. Correspondingly, the download times increase from $\approx$ 150\,s to  $\approx$ 250\,s.

\subsubsection{Device Heterogeneity: A Single Point\\ of Weakness}

\begin{figure}
  \subfigure[Memory Consumption]{\includegraphics[width=\columnwidth]{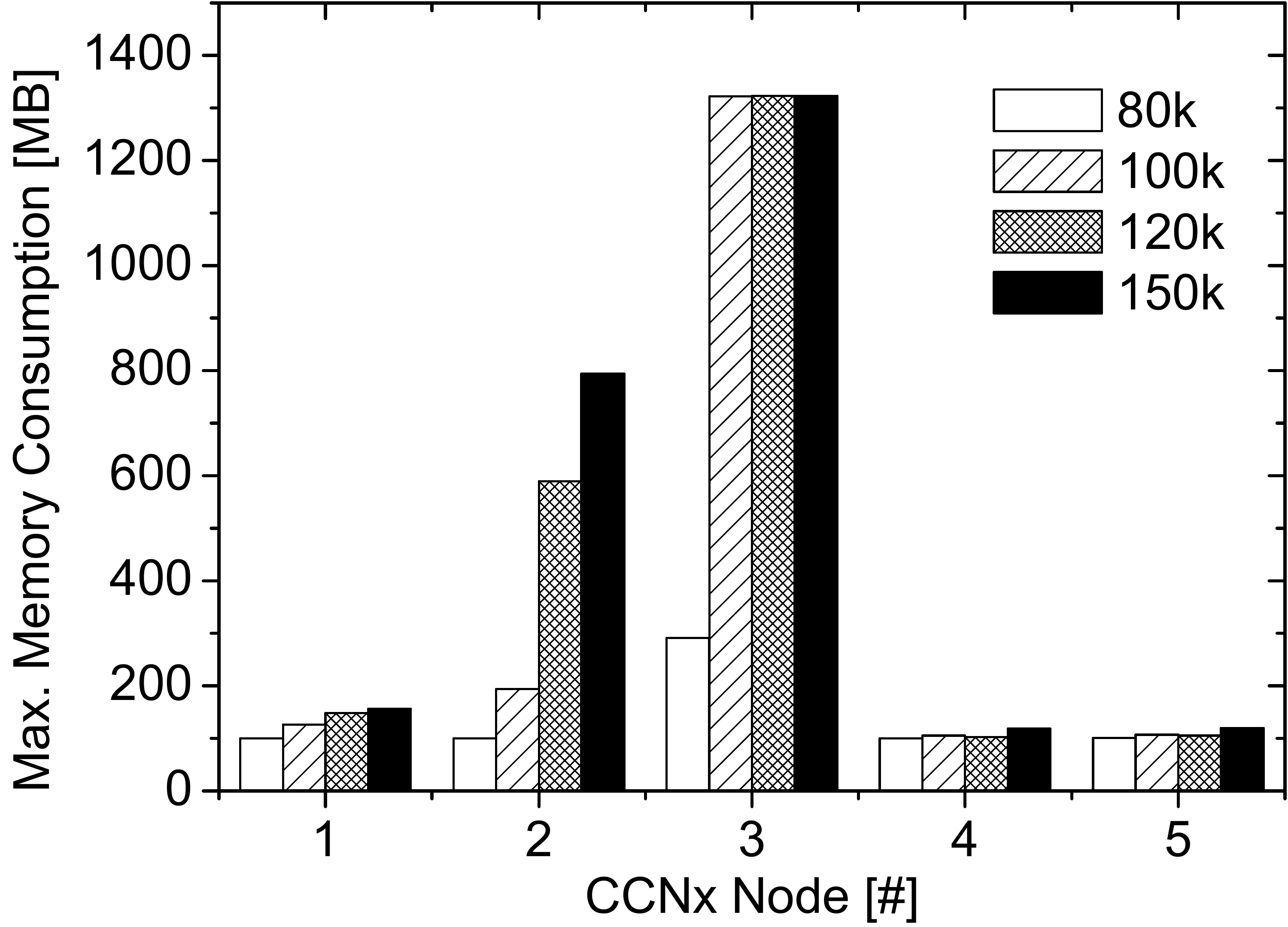}\label{fig:noContent5hopsMemory}}
  \subfigure[Average CPU Load]{\includegraphics[width=\columnwidth]{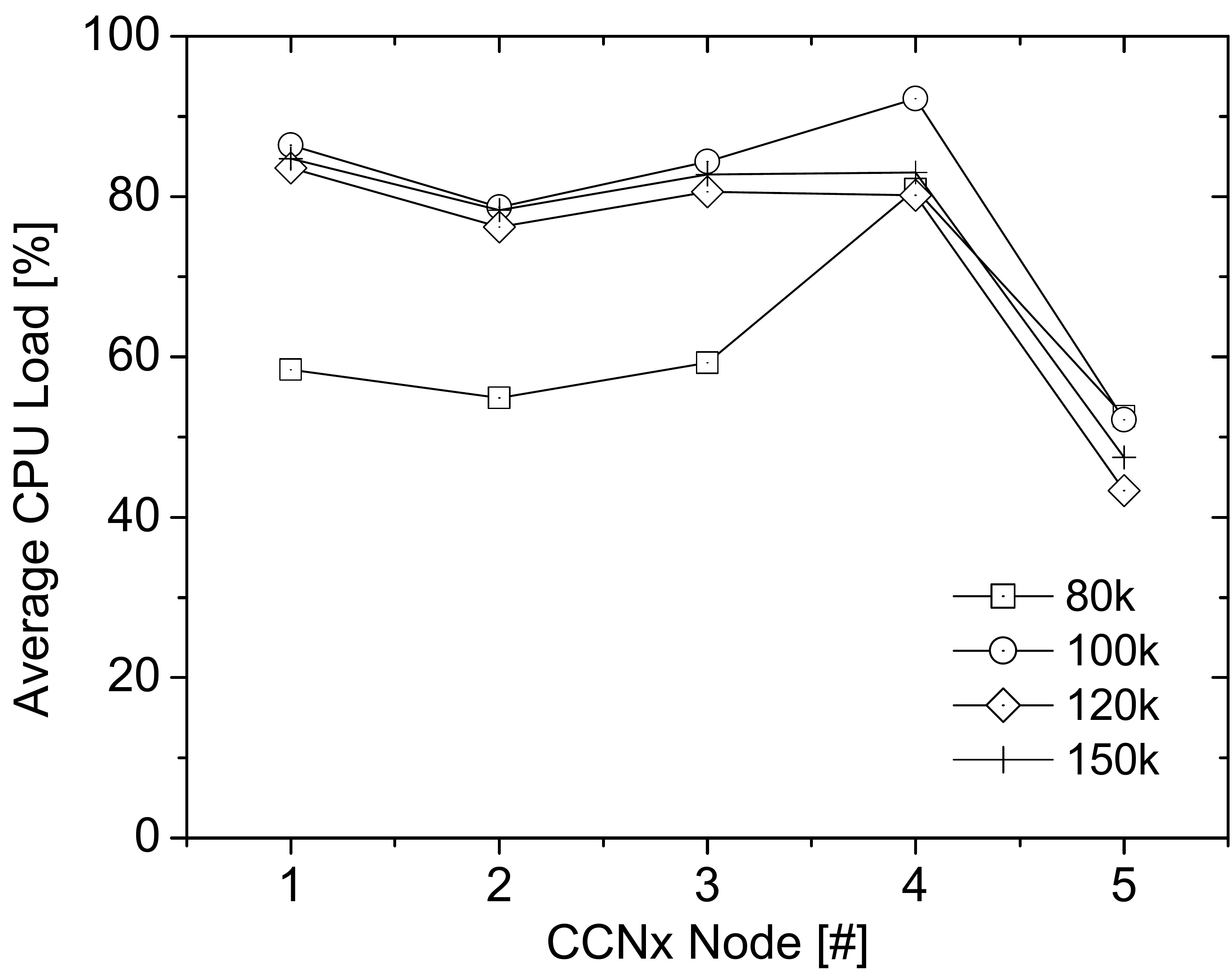}\label{fig:noContent5hopsCpu}}
  \caption{Load per hop for a chain of 5 routers while initiating a 80k,
  100k, 120k, and 150k different Interests for non-existing
  content}\label{fig:noContent5hops}
\end{figure}

It is a valid assumption that the content distribution system will consist
of heterogeneous devices in terms of all performance metrics. In this second experiment, 
 we introduce device heterogeneity by weakening a single router, the 4th hop (CCNx4), in a controlled way. We want to study the reaction of state management and network performance to this well-defined degradation. 

For an initial observation of the dependency on the weakest node, we reduce the CPU capacity of CCNx4 to 25\,\% (600~MHz) and recap the scenario from
Section~\ref{sec:resourceex} for 80k, 100k, 120k, and 150k subscriptions of
non-existing content. Independent of the capacity of the network infrastructure, the consumer
initiates content subscriptions and continuously refreshes its Interests,
which then  propagate towards the content repository.

Figure~\ref{fig:noContent5hops} shows the maximal
memory consumption and the average CPU load per hop during the measurement
period. It is clearly visible that the required memory mainly depends on
the position of the node within the topology. Memory requirements on the 
 single path fluctuate by two orders of magnitude. The predecessor of the node
with the lowest processing capacities (i.e., the 4th~hop) needs 50\% --
500\% more memory than any other nodes. 

In this scenario of
reliable transport for signaling messages, a bottleneck node puts pronounced load onto its
predecessor as the forwarding of the refresh messages retain. The reason for this strongly unbalanced performance picture can be found in RTT fluctuations: The weak on-path router appears like a periodically effective blockade for Interest transmission.  In general, heterogeneous network conditions lead to a discontinuous propagation of Interest states and thus to a 
fluctuating performance. The successors of the weak router then do not receive state updates in time, and the predecessors experience large waiting periods. Correspondingly, the results displayed in Figure~\ref{fig:noContent5hops} impressively demonstrate the effects of time fluctuations in content-centric routing.

\subsubsection{Steady-State Stability}

In the next step of our analysis, we take a closer look on gradual effects of routing heterogeneity. We observe corrective mechanisms of the network (i.e., Interest retransmissions) depending on router asymmetry. Interest retransmissions serve as the key indicator for timeouts due to router overload. For this task, we configure CCNx4 with four different processing capacities related to the other CCNx routers: 2,400~MHz (homogeneous), 1,200~MHz (50\,\% capacity), and 600~MHz (25\,\% capacity).   

\begin{figure}
  \includegraphics[width=\columnwidth]{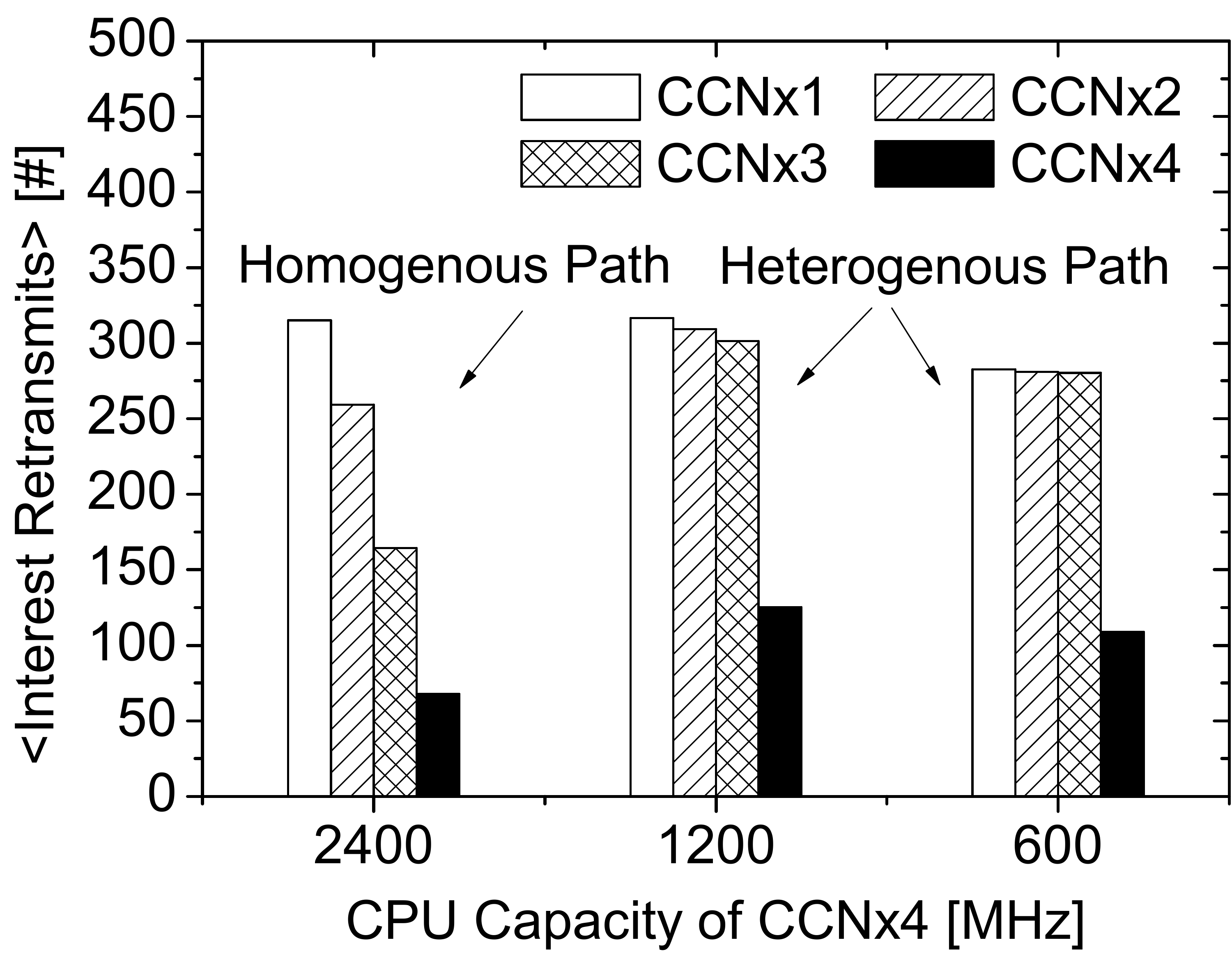}
  \caption{The effect of router strengths on Interest trading}\label{fig:BN_x_retrans}
\end{figure}

Surprising results are shown in Figure~\ref{fig:BN_x_retrans}. Evidently we see that network behavior switches on the occurrence of router heterogeneity, independent of its strengths. The characteristic behaviour of the balanced network is a steady decay of Interest retransmits towards the source, as data delivery gets faster and more reliable in proximity to the publisher. However, at the first occasion of a `bottleneck', the picture flips. Interest retransmission drastically increases and all routers except for the bottleneck equally see about the maximal rate of retransmissions in this scenario. State retransmissions at the weak forwarder (CCNx4) instantaneously doubles to the level of managed states this router can cope with.

This experiment clearly shows how sensitive content-centric routing reacts to varying network resources. A light disturbance of the state propagation process reveals the instability of a steady-state flow by immediately turning content transport into a significantly different condition of maximal error management.

\subsubsection{Complex Inhomogeneities}

\begin{figure*}[ht]
  \subfigure[Pending Interests]{\includegraphics[width=0.33\textwidth]{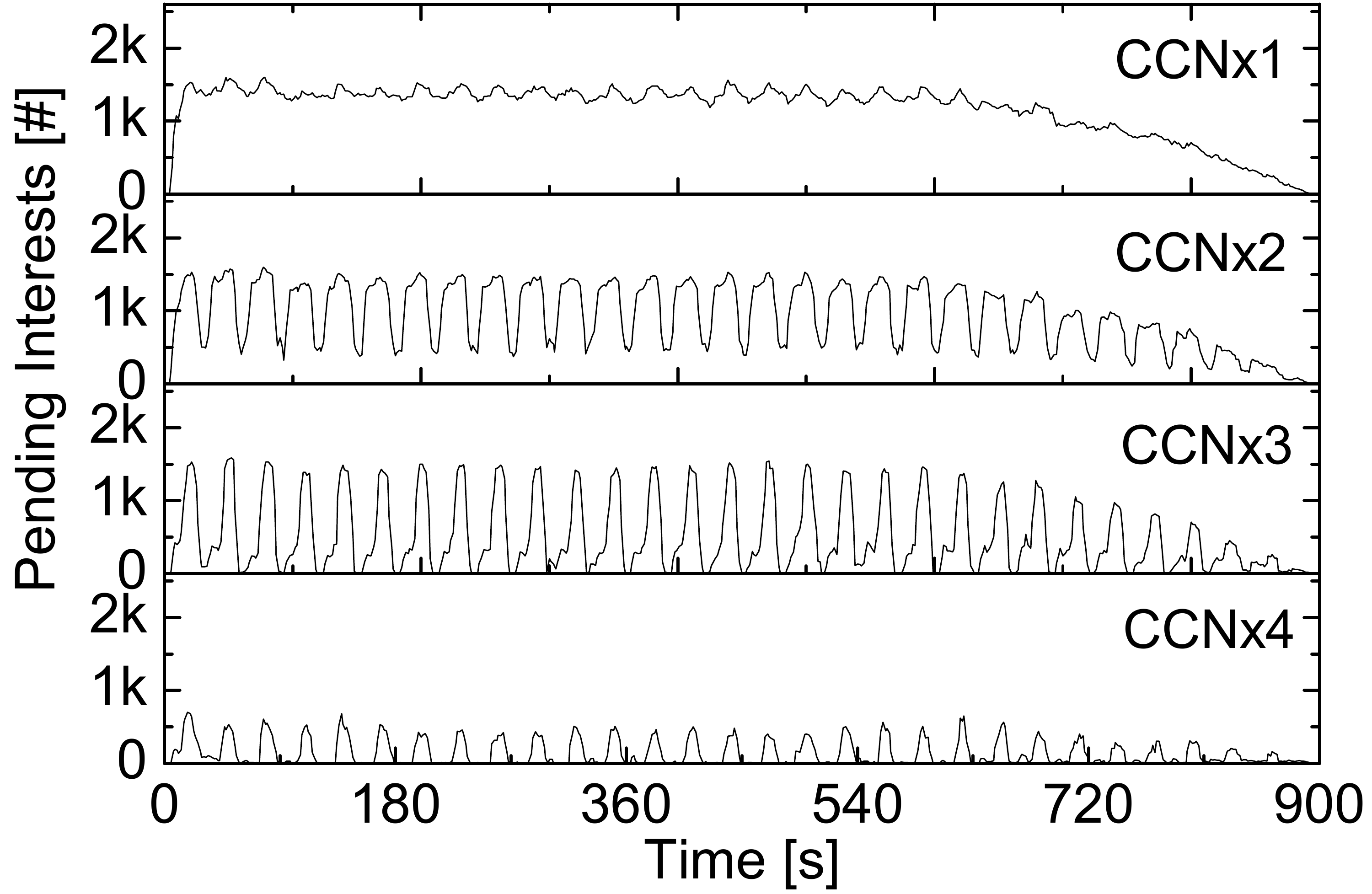}\label{fig:anp_chain_pi}}
  \subfigure[Interest Retransmits]{\includegraphics[width=0.33\textwidth]{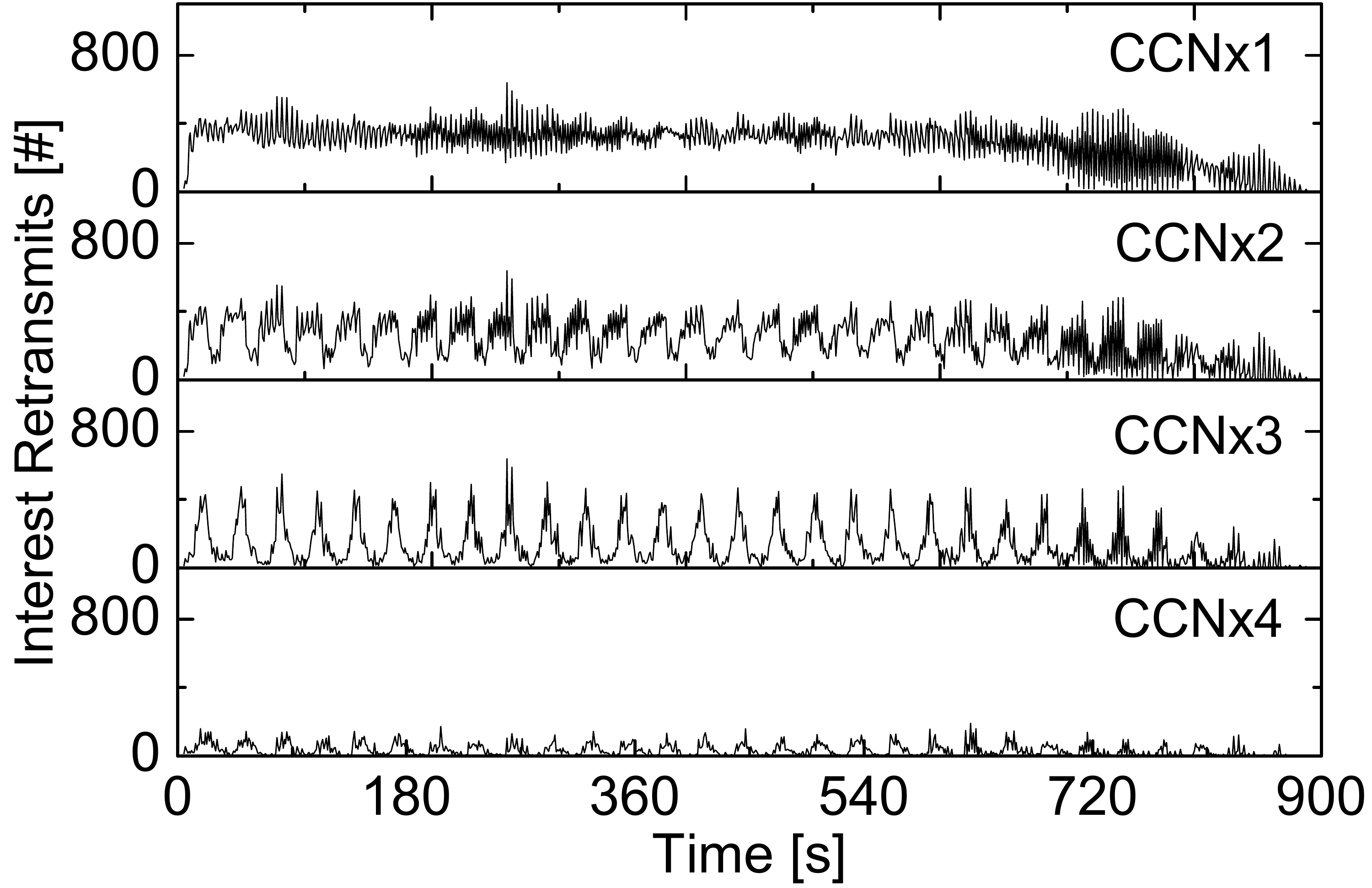}\label{fig:anp_chain_re}}
  \subfigure[Network Utilization]{\includegraphics[width=0.33\textwidth]{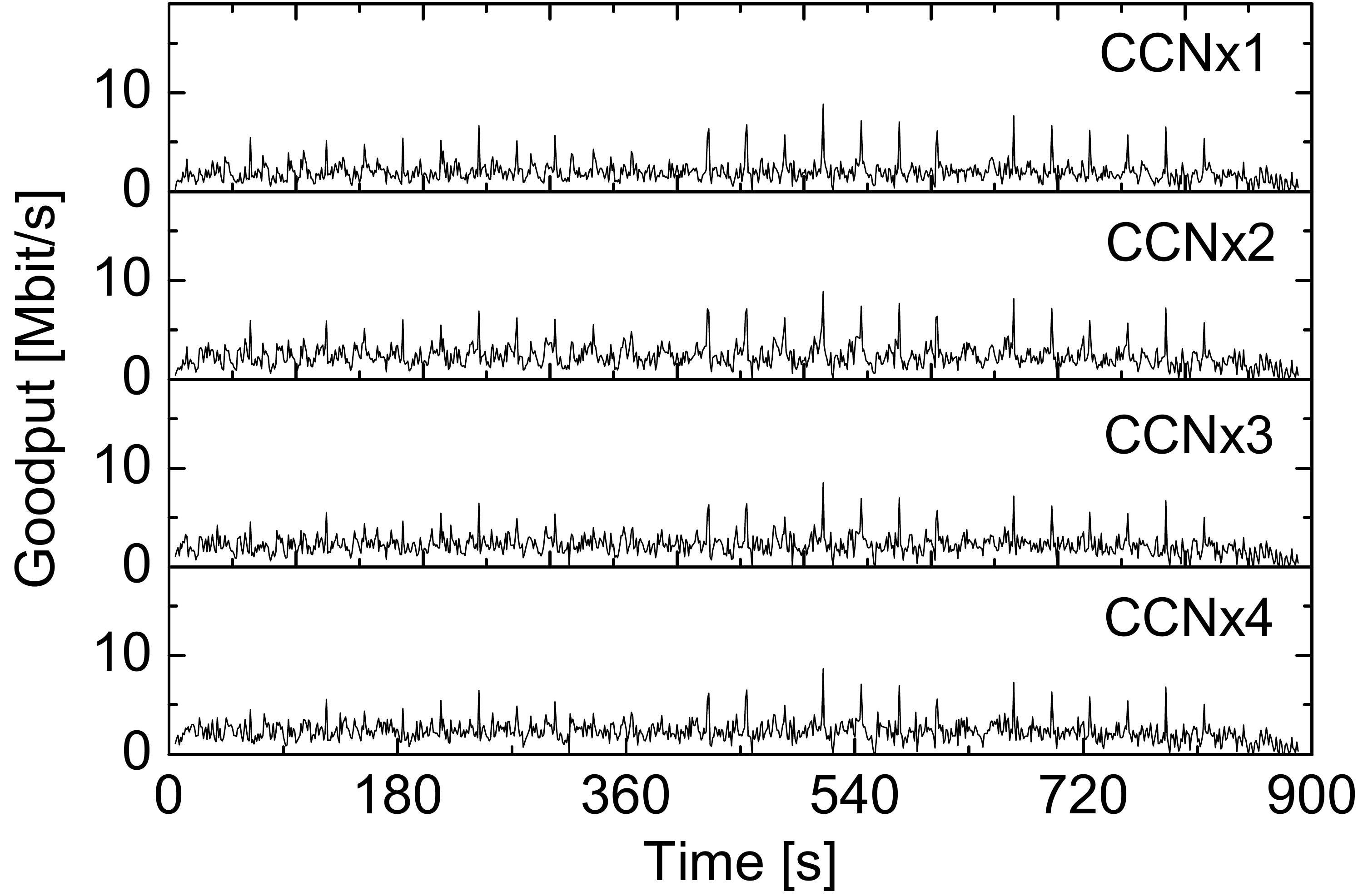}\label{fig:anp_chain_gp}}
\caption{Routing and forwarding performance in a five-hop network with alternating CPU reductions} \label{fig:anp_chain}
\end{figure*}

In our final experiment concerned with content routing, we want to explore situations of largely decorrelated network conditions. Therefore we configure all routers to admit fast changing resources occurring in anti-cycles. In detail, each intermediate router (CCNx2, \dots, CCNx4) is forced into a periodic CPU reduction by 90\,\% of 30\,s. Resource reduction periods were shifted between routers by 15\,s so that at least one of the three routers in the forwarding chain was kept in challenged conditions. The objective of this repelling setup, which similarly may well occur from different side traffics in a meshed backbone, is to analyse the vulnerability of hop-by-hop state maintenance in content-centric routing.

Results of this alternating resource scenario are displayed in Figure~\ref{fig:anp_chain}. Visualisation is completely analogous to the initial view on a homogeneous network (see Figure~\ref{fig:hom_chain}). The course of pending Interests as well as Interest retransmissions open a distinguished view on the fine-grained sensitivity of content routing to neighboring router conditions. State provisioning fluctuates on the resource resolution scale of 30\,s throughout the network. More importantly, data transmission rates drop down to about 2.4\,Mbit/s,  while the overall load of Interest states remains compatible to the homogeneous network. Uncoordinated network resource availability thus leads to a low overall performance in conjunction with high network resource consumptions. Time-to-completion for each single file download correspondingly explodes to 900\,s for the same 10\,Mbit files as in our initial experiment. It should be recalled that network capacities do allow for a simultaneous download of all 500 files within 10\,s.

\subsubsection{A Comparative Summary}

In the different scenarios of our experimentally-driven analysis, we have revealed a number of structural weaknesses and effects immanent to the current state of content-centric routing in heterogeneous multi-hop scenarios. All examples lead to severe service degradations. Nevertheless the question arises on how these insights combine to a consistent view.

\begin{figure*}
  \subfigure[Homogeneous Network]{\includegraphics[width=0.33\textwidth]{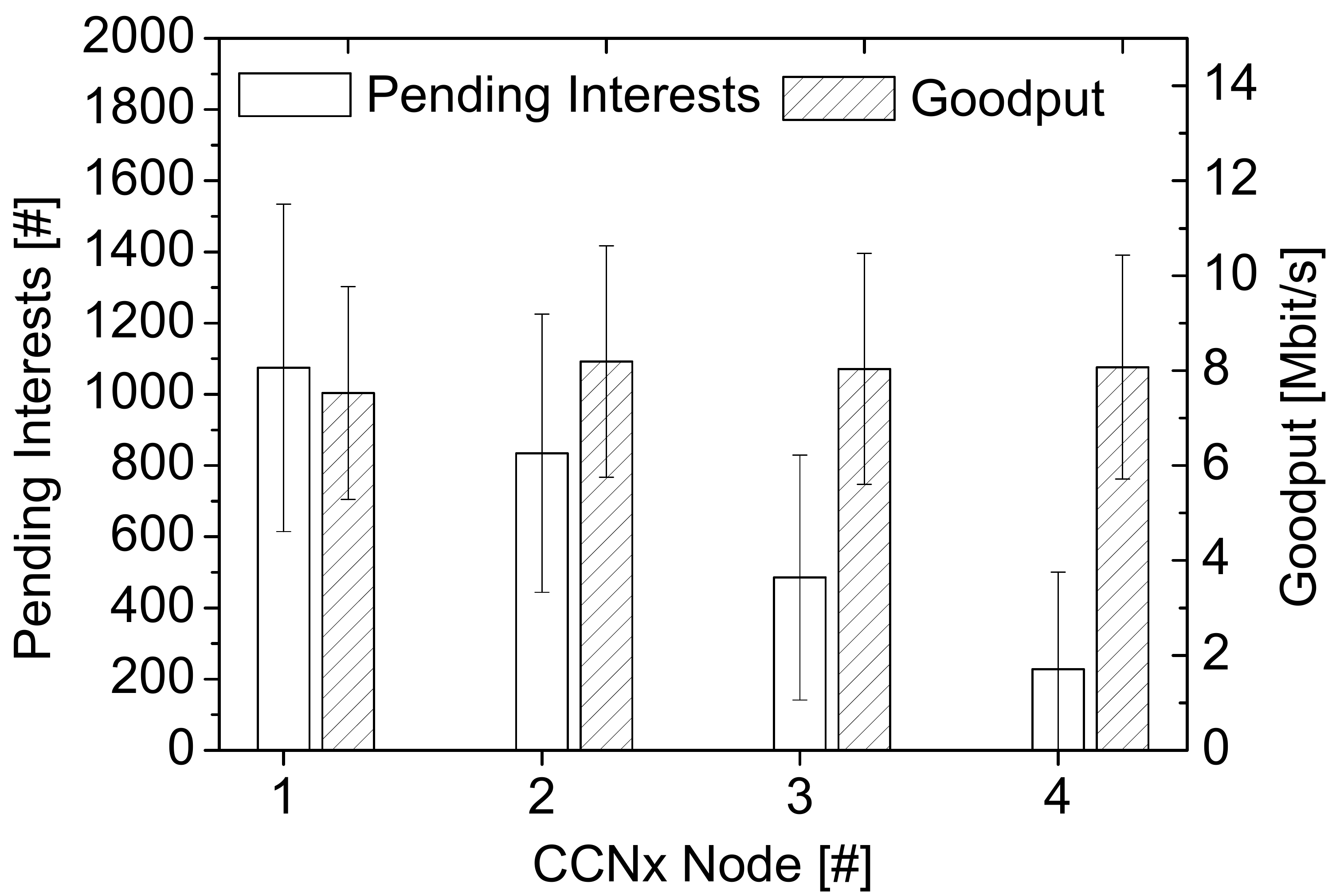}\label{fig:him_chain_pidgood}}
  \subfigure[Single Point of Weakness]{\includegraphics[width=0.33\textwidth]{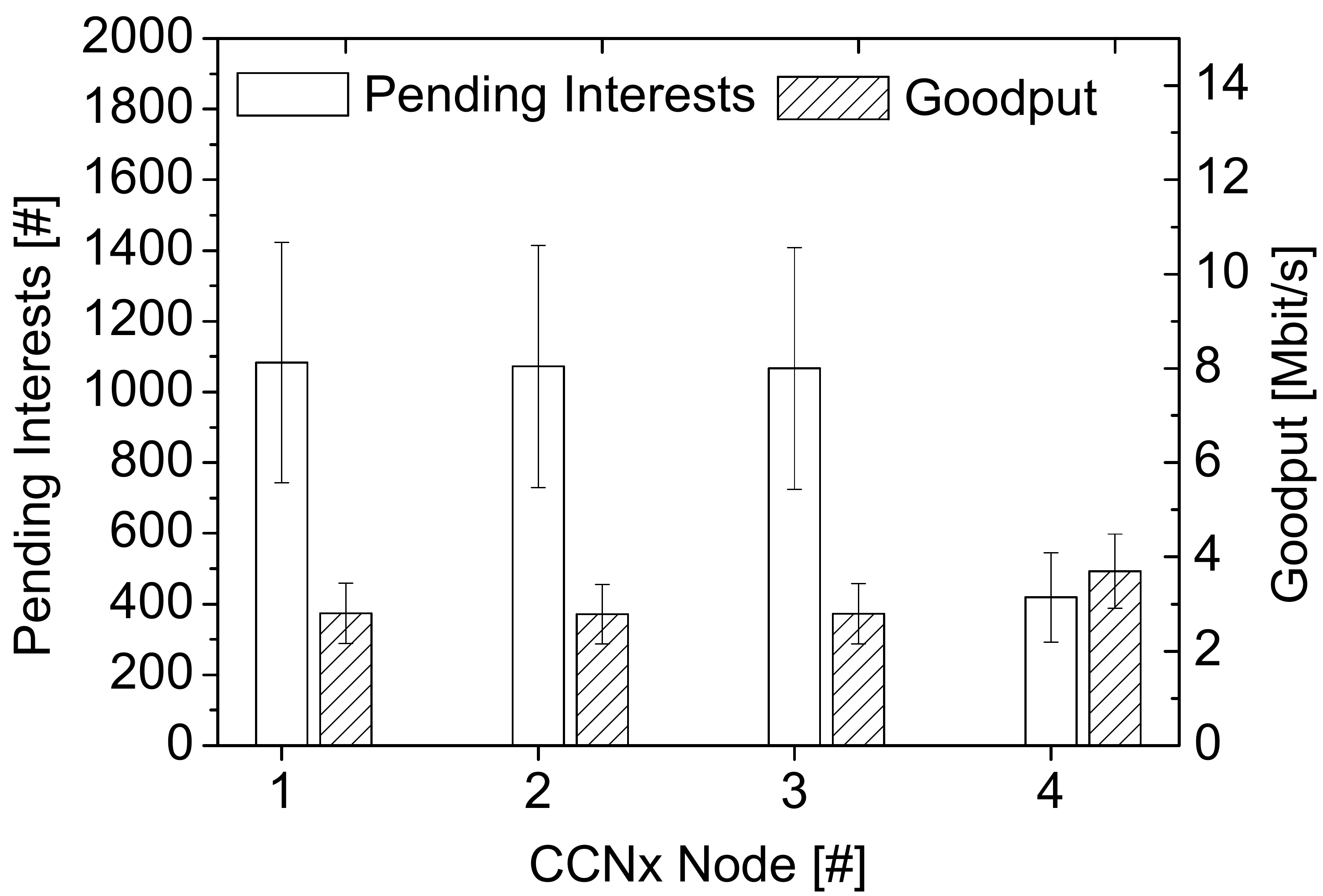}\label{fig:bn4_chain_pidgood}}
  \subfigure[Alternating Resources]{\includegraphics[width=0.33\textwidth]{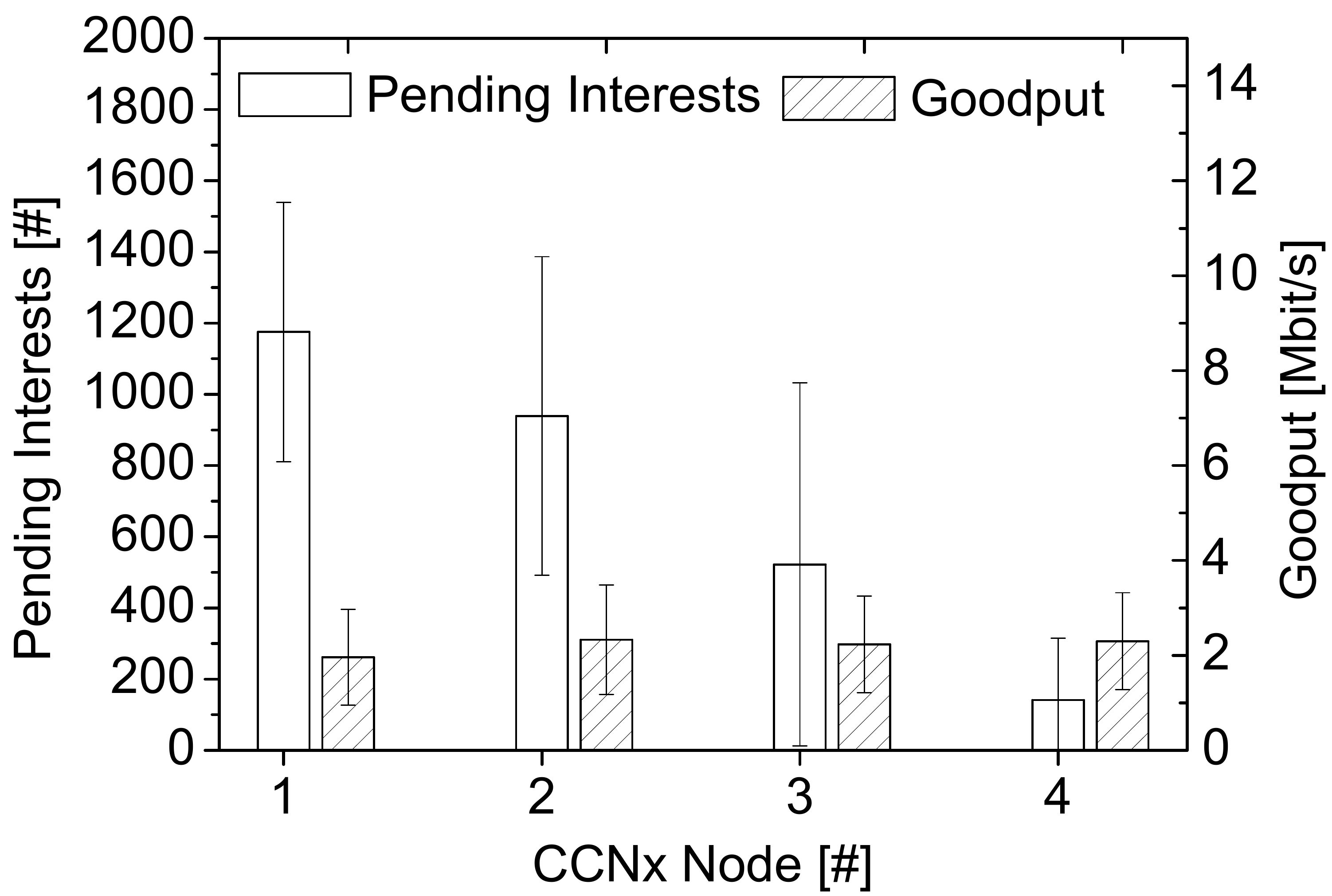}\label{fig:anp_chain_pidgood}}
\caption{Comparison of state management and forwarding performance in different network scenarios (mean and standard variation)} \label{fig:chain_pidgood}
\end{figure*}

We will try to present an overall picture in the following. Figure \ref{fig:chain_pidgood} contrasts the load imposed onto the infrastructure by Interest states with the average network performance in the three experimental scenarios, homogeneous network, single point of weakness, and alternating resources at routers. The striking picture in all three settings is that the efficiency of network utilization is low on the overall, but drastically drops whenever inhomogeneities occur. The hop-by-hop forwarding performance thus must be considered rather fragile. In contrast, network state propagation attains various patterns, but always remains at compatible level at the router of maximal load. 

These observations suggest the following rule of thumb for CCN routing performance: State maintenance always follows the maximal requirements, while forwarding performance will adapt to the weakest resource in place. This overall picture is clearly inefficient and future work on ICN solutions would largely benefit from improving this behaviour.

\section{Examples of Attack Scenarios}\label{sec:attacks}

In this section, we will outline and briefly discuss new attack scenarios that arise from data-driven state management in ICN. We will illustrate these examples by terms related to our experiments on CCNx (e.g., PIT), but should emphasize the general applicability to subscription resp.~publication states as discussed in Section~\ref{sec:ps}.

\subsection{Attacks Related to Resource Exhaustion} \label{sec:attacksresources}

As shown in the previous section, routing and forwarding capacities of the infrastructure can be easily compromised by overloading its content request or interest tables. As  non-aggregable name requests for locally unavailable content propagate through the network, resource exhaustion attacks can be transparently initiated from the remote. For this purpose it remains completely indifferent whether hardware resources are drained at unrestricted PIT sizes or table space exhausts according to various limiting configurations. Correspondingly, FIB overflows at routers occur in response to excessive publications or updates of names. Details of the attacker's effects depend on the state-dropping strategy---for simplicity we assume dropping tails as used by CCNx.
In addition, virtual resources may also be depleted. The injection of bogus Interests disturbs ICN flow control mechanisms at routers, for example, because it reduces request limits.

\textbf{Remotely Initiated Overload}
\quad
An attacker that controls one or more machines (a botnet) may initiate massive requests for locally unavailable content. Corresponding interests propagate towards the publisher and eventually accumulate at some content router causing overload conditions. Depending on its intensity, this attack will lead to a service impairment or DoS for the (remote) content distribution tree(s) branching at the degraded router, unless the networking system is able to re-route the requests. It is worth noting that timeouts at regular users on the subtree will initiate retransmission `storms' and thereby amplify the attack.

\textbf{Piling Requests due to a Slow Source}
\quad
Performance of a content source may be degraded by artificially high numbers of direct requests causing slowed down responsiveness. Alternatively, a captured source or its overloaded access router may drastically increase response times of content delivery. In slowing down a (popular) source, an attacker lowers the data return rate and thereby the extinction of pending interests at all routers on paths to receivers. Thus attacking a single point may result in a widely increased load at the routing infrastructure.

\textbf{Mobile Blockade}
\quad
A mobile node (MN) may issue a large number of invalid interests that block the PIT of the mobile access router for the period of state timeout. In a shared link-layer environment that cannot easily detect its departure, the mobile adversary can traverse neighboring networks on circular routes and continue to offload its interest bundle with the effect of a  blockade of the regionally available networks. Initial countermeasures are difficult to apply, as the retransmission of interests is part of the regular mobility pattern in ICN.

\textbf{Fooling Rate Limiting}
\quad
Current ICN approaches~\cite{yamwz-csfp-12} propose rate limiting to restrict the number of Interest states. 
Its main purpose is flow control to avoid congestion. In contrast to common believes (see \cite{gtuz-ddnn-12} for discussions) we argue that this is not an appropriate countermeasure to protect the ICN distribution system against attacks.
An attacker can easily create an interest storm that exceeds the anticipated interest limit. The dedicated router will throttle the number of accepted interests per interface or interface+prefix, and finally ignore subsequent interests. Consequently, a single end user blocks a prefix or harms all members of its domain. 

Note, applying rate limiting per end host (or user) is non-trivial in ICN. ICN explicitly discontinues the concept of host identifiers (e.g., due to security reasons). Thus, a router cannot track particular sources that send an unexpected amount of interests. Even if routers are enabled with such a function, an attacker can spoof addresses.

The same holds for push-back mechanisms \cite{gtuz-ddnn-12}, which signal an overload towards the source and thus try to isolate an attacker. In addition, an attacker that would receive such a control message can ignore it.

\subsection{Attacks Related to State Decorrelation} \label{sec:attacksdecorrelation}

ICN requires consistent states during the request routing phase {\em and} the asynchronous content delivery. While bogus announcements or flapping of routes may introduce loops or increase the likelihood thereof, incoherent forwarding paths may result in partial content transmission that uses network resources without success in data delivery.

\textbf{Infringing Content States}
\quad
An attacker that controls end systems or content routers could announce updates of content or cache appearances at a frequency that exceeds the (local) content request routing convergence time. As a consequence, the overloaded announcement or mapping system will be unable to correctly process the updates of proper content sources or caches with the effect of incomplete content representation and erroneous data replication states. Content requesters will be thus led into false retrievals or access failures. As content announcement is commonly built on soft-state approaches, failures will eventually timeout after a period of undisclosed inconsistency, which the adversary could initiate in a momentary attack.

\textbf{Timeout Attack}
\quad
The timeouts of pending interests fire independently in a distributed network environment, which is normally healed by early refreshes. An attacker that controls one or more machines may interfere with state coordination by degrading the performance of two or more on-path routers so that request routing and data forwarding exhibit large, fluctuating delays in the order of the timeout. As a consequence, intermediate timers will erase interests with large probability prior to data delivery or retransmission requests. Corresponding receivers will suffer from DoS for the affected chunks or arbitrarily large transmission delays.

\textbf{Jamming Attack}
\quad
A node on a shared link may issue a large number of content requests without maintaining the interests at its own (loosing interest). Content will then arrive at the local link without a receiver. This scenario is particularly harmful in mobile environments of limited bandwidth. A mobile attacker can jam a region by traversing shared radio links while requesting bulk data.   

\subsection{Attacks Related to Path and Name Infiltration} \label{sec:attackinfiltration}

ICN raises content names and cache locations to first class objects and must therefore remain open to naming and placing data. The request routing system carries routes to names in its FIB or a mapping service, both of which are vulnerable to resource exhaustion and route poisoning. While an explosion in the pure number of names may be mitigated in parts by aggregation according to some authoritative  naming conventions like in today's domain names, bogus route infiltration must be considered the more delicate issue.

\textbf{Route Hijacking}
\quad
An adversary may announce (or register) routes to cached copies of any content object. Content requests from its vicinity are then directed towards the malicious system and -- if unanswered or retarded -- lead to long-lasting forwarding states and a possible DoS. This threat can be mitigated by resource-intensive attempts to route towards multiple locations that become increasingly painful when an attacker controls a botnet and injects invalid routes at large scale.

\textbf{Route Interception}
\quad
Alternatively, an adversary may keep record of the valid routes to the content, while it announces various malicious routes. Content requests can then be intercepted and forwarded to the proper location. The intruder will thus gain knowledge of the content consumed in the networks under attack, while receivers experience a normal network behavior. This privacy violation may allow for highly selective information, whenever the attacker is topologically close to the victim.




\section{Conclusions \& Discussions}\label{sec:discussion}

In this paper, we have analyzed network instabilities in
information-centric networks that are caused by (a) data-driven state
management and (b) backbone states initiated by end users.

Some threats are easy to anticipate (e.g., resource exhaustion), others
are more intricate due to the complex interplay of distributed states
(e.g., state decorrelation). For the latter previous practical
insights in the design of (conceptually related) multicast protocols
already revealed good and bad design options. One of the major design
goals of Bidirectional~PIM~\cite{rfc-5015}, for example, was "`eliminating
the requirement for data-driven protocol events"'---after the operating
experiences with data-driven DVMRP or PIM-SM. We are somewhat puzzled that
those insights seem to have been ignored by the ICN community. With this
paper we want to stimulate the discussion about basic threats and attacks
on content-centric backbone routing. We admit that resolving the identified
problems is definitely challenging as this either reduces capabilities of
the ICN infrastructure or flexibility at the end user.

The exhaustion of  memory and processing resources following excessive
state allocations and manipulations was identified as one major reason for
service degradation and DoS attacks. An obvious approach to  mitigate the
resource exhaustion problem is to limit the rates of state injection into
the network. Applying restrictions per domain (e.g., \cite{yawzz-afndn-12})
will create new attack vectors from the interior, while limiting users 
will require tracking of end nodes and leads to traffic shaping and bandwidth restrictions. As content states will accumulate in the network, and inter-provider deployment almost surely will lead to a heterogeneous, unbalanced design, rate limiting may milden, but cannot effectively prevent the resource exhaustion problems discussed in this paper.

The other major threat to ICN stability arises from the ease of malicious name or route infiltration. Even though similar vulnerabilities are known from BGP, ICN backbones will be populated by data-centric states---created, modified or deleted by any end-user of the network---and thus largely magnify the problem scope of poisoning the control plane. 

 Current CDN deployments remain agnostic of these infringements by running under proprietary regimes.
Still, if we want the Future Internet to remain open for every content published by any user, we cannot impose restrictions that approach today's CDN regulations in any way.

\bibliographystyle{IEEEtran}
\bibliography{/bib/own,/bib/rfcs,/bib/ids,/bib/theory,/bib/layer2,/bib/internet,/bib/transport,/bib/overlay,/bib/vcoip,/bib/ngi,/bib/security}
\normalfont

\end{document}